\documentstyle[prl,aps,epsf]{revtex}
\newcommand{\BEQ}{\begin{equation}}
\newcommand{\EEQ}{\end{equation}}
\newcommand{\BEA}{\begin{eqnarray}}
\newcommand{\EEA}{\end{eqnarray}}
\renewcommand{\H}{{\cal H}}
\renewcommand{\O}{{\cal O}}
\renewcommand{\d}{{\rm d }}
\newcommand{\dx}{{\rm d}x\,}
\renewcommand{\S}{S_{\rm ep}}
\newcommand{\p}{\partial}
\newcommand{\Q}{R}
\newcommand{\erfc}{{\rm erfc}}
\newcommand{\dy}{{\rm d}y\,}
\newcommand{\dt}{\frac{1}{N}}
\newcommand{\e}{{\rm \varepsilon }}
\newcommand{\half}{\frac{1}{2}}
\newcommand{\tr}{{\rm tr}}
\newcommand{\minfty}{{-\infty}}

\newcommand{\I}{{\cal I}}
\newcommand{\xav}{<\!x\!>}

\newcommand{\nn}{\nonumber \\}
\newcommand{\y}{{\overline y}}
\def\dbarrm {{\mathchar'26\mkern-11mu{\rm d}}}                         %
\newcommand{\fix}{{\Bigl.\Bigr|}}
\renewcommand{\thesection}{\arabic{section}}
\draft
\begin{document}
\title{Thermodynamic picture of the glassy state
gained from exactly solvable models}
\author{
Th.~M.~Nieuwenhuizen}
\address{Department of Physics and Astronomy, University of Amsterdam
\\ Valckenierstraat 65, 1018 XE Amsterdam, The Netherlands}
\date{Version: June 29, 1998; printout: \today}
\maketitle
\begin{abstract}
A picture for thermodynamics of the glassy state was introduced
recently by us 
(Phys. Rev. Lett. {\bf 79}  (1997) 1317; {\bf 80} (1998) 5580).
It starts by assuming that one extra parameter, 
the effective temperature, is needed to
describe the glassy state. This approach connects responses of
macroscopic observables to a field change with
their temporal fluctuations, and with 
the fluctuation-dissipation relation, in a generalized,
non-equilibrium way. Similar universal relations do not hold between
energy fluctuations and the specific heat.

In the present paper the underlying arguments are discussed
in greater length.
The main part of the paper involves details of the exact dynamical
solution of two simple models introduced recently: uncoupled
harmonic oscillators subject to parallel Monte Carlo dynamics, and
independent spherical spins in a random field with such dynamics.
At low temperature the relaxation time of both models 
diverges as an Arrhenius law, which causes glassy behavior
in typical situations.
In the glassy regime we are able to verify the 
above mentioned relations for the thermodynamics of the glassy state.

In the course of the analysis it is argued that stretched exponential
behavior is not a fundamental property of the glassy state, though
it may be useful for fitting in a limited parameter regime.

\end{abstract}

\section{Introduction}
\setcounter{equation}{0}\setcounter{figure}{0} 
\renewcommand{\thesection}{\arabic{section}.}

Thermodynamics is an old but very powerful subject. It applies to
a wide variety of systems ranging from ideal gases to crystals 
and black holes. Important contributions to its development were made by
Carnot, Clausius, Kelvin, and Boltzmann.  
Equilibrium thermodynamics, better called ``thermostatics'',
is a well understood subject, and applied every day in many fields 
of science.  The work of Gibbs showed its tremendous generality
via its relation to statistical physics (i.e. partition sums). 
We  shall explain, however, that precisely this success has been a 
stand in the way for the systems our interest: glasses.

Non-equilibrium thermodynamics 
for systems close to equilibrium was worked out in the first half of
this century. Typical applications are systems with heat flows,
electrical currents, and chemical reactions.
The basic assumption is the presence of local thermodynamical equilibrium,
and the basic task is to calculate the entropy production.
Important contributions to this field were made by de Donder, 
Prigogine, de Groot and Mazur.

Non-equilibrium thermodynamics for systems far from equilibrium
has long been a field of confusion. A typical application is window glass.
Such a system is far from equilibrium: a cubic micron of glass is neither
a crystal nor an ordinary undercooled liquid. 
It is an undercooled liquid that, in the glass formation process,
 has fallen out of its own metastable equilibrium. The glassy state is
inherently a non-equilibrium state: a substance that is a glass in
daily life (time scale of years) would behave as a liquid on
geological time scales. If each 500 years a picture
would be taken of a window glass, then the movie composed of these
pictures would look very much like  a movie of a soap film.
  
Until our recent works on this field, the general consensus reached 
after more than half a century of research was: 
{\it Thermodynamics does not work for glasses,
because there is no equilibrium}. Even before going into any detail,
it is clear that this conclusion itself is confusing, because
{\it  thermodynamics should also apply outside equilibrium}. 
Inspired by the success of Gibbsian theory, the whole
non-equilibrium part of thermodynamics had been forgotten!
The correct formulation should of course have been:
{\it Equilibrium thermodynamics does not work for glasses,
because there is no equilibrium}, 
surely a less surprising  and non-embarrassing statement. 
(This history shows once more how regretful it is that equilibrium
thermodynamics did not get known under its most proper 
name,  ``thermostatics''.)

The negative conclusion about the applicability of thermodynamics
was mainly based on the failure to understand the Ehrenfest relations 
and the Prigogine-Defay ratio.  It should be kept in mind
that, so far, the approaches leaned very much on equilibrium ideas.
Well known examples are the 1958 Gibbs-DiMarzio
{}~\cite{GibbsDiMarzio} and the 1965 Adam-Gibbs~\cite{AdamGibbs} papers,
while a 1981 paper by DiMarzio has title ``Equilibrium theory of
glasses'' and a subsection ``An equilibrium theory of glasses is
absolutely necessary''~\cite{DiMarzio1981}.
In our opinion such approaches are not applicable, due to the
inherent non-equilibrium character of the glassy state.
In the course of the present work we shall encounter lots of 
instances where such approaches indeed fail to describe the physics.
Notice, however, that this  immediately rules out the by far most 
discussed model glass, namely the Gibbs-DiMarzio theory~\cite{GibbsDiMarzio},
as a viable model for a realistic glass. For instance, it would
predict the original Ehrenfest relations to be always satisfied, 
in violence with experiments to be discussed.

In our view the current lack of a thermodynamic description is 
quite unsatisfactory, since so many decades in time are involved, 
ranging from the microscopic sub-picosecond regime
to, for silicate-rich glasses, almost the age of the solar system, 
thus covering more than 25 decades. Naively we expect that
each decade has its own dynamics, basically independent of the
other ones. We shall find support for this point 
in the models that we shall investigate below.

Near the glass transition a glass forming liquid exhibits
smeared discontinuities in quantities such as the heat capacity, the
expansivity and the compressibility. This look similar to 
continuous phase transitions of the classical type, i.e., with specific heat
exponent $\alpha=0$, even though the analogy is not perfect, 
because the smaller specific heat value occurs below the glass transition.
It was then investigated whether the
jumps satisfy the two Ehrenfest relations (the analogs for second order
transitions of the Clausius-Clapeyron relation of a first order transition).
As reviewed recently~\cite{Angell}, it was found that the first
Ehrenfest relation, involving the jump in the compressibility, is always 
violated, while the second one, involving the jump in the specific
heat, is usually satisfied, but not always. It has become fashionable
to combine these two relations by introducing the so-called
Prigogine-Defay ratio $\Pi$. For equilibrium transitions this quantity
should be equal to unity, and it was generally expected that it cannot take
values below unity. In glasses typical values are said to lie in the range 
$2<\Pi<5$, even though a very careful experiment on polystyrene
led to $\Pi\approx 1$ ~\cite{RehageOels}.

Our recent studies have radically changed the view point.
We have realized that
the first Ehrenfest relation is automatically satisfied, the only
subtlety being its proper interpretation. We have also put forward
that the Maxwell relation and the second Ehrenfest relation are
modified in the glassy state, due to lack of equilibrium~\cite{NEhren}.

We have investigated the possibility that, within a yet 
unknown class of systems, 
the glassy state is described by one extra state variable.
This is basically the age of the system, or the cooling rate under
which the glass has been formed. We realized that in thermodynamics
the {\it effective temperature} $T_e$ is a more useful extra 
parameter~\cite{Nthermo}\cite{Nhammer}.

This approach has led to a picture for thermodynamic 
relations between values of macroscopic observables
~\cite{Nthermo}\cite{NEhren}
Later it was extended to their fluctuations~\cite{Nhammer}. 
The picture also incorporates 
the so-called fluctuation-dissipation relation (FDR), put forward
in works by Horner~\cite{Horner1}\cite{Horner2}\cite{CHS}
 and Cugliandolo and Kurchan
~\cite{CuKu}; for a review, see ~\cite{BCKM}. This relation has
become a central point in research on off-equilibrium systems.
Our more general approach shows that the effective
temperature that occurs in thermodynamics and the one that 
occurs in the fluctuation-dissipation relation are almost identical.

In the course of our work
we gained insights from analytical results combined with
educated guessing on the $p$-spin interaction spin glass.
Some initial studies had the purpose to find 
the physical meaning of the non-equilibrium
replica free energy in spin glass models~\cite {Nmaxmin}.
It has turned out that replica theory provides the two-temperature
off-equilibrium free energy that we shall discuss in much more 
general context~\cite{Nthermo}.  
The basic drawback of the $p$-spin model is that 
dynamics is not solved in the activated regime.
For a model of directed polymers in a correlated random potential
the situation is a little better, but so far it 
also lacks a complete solution in the activated regime~\cite{Ndirpol}.
Another model is the backgammon model, for which the dynamics at zero
field has been partly solved~\cite{FranzRitort}\cite{GodrecheLuck}. 
One could couple the system to a particle bath, and the chemical potential 
would play the role of an external field. So far this case remains to
be worked out.

More promising is a model of independent harmonic oscillators
with parallel Monte Carlo dynamics, introduced recently by 
Bonilla, Padilla and Ritort~\cite{BPR}.
For this model the Hamiltonian and thus the statics is trivial.
Nevertheless, the  exactly solvable dynamics exhibits interesting glassy
aspects. Since it has a too simple behavior in a field, we have
recently studied a related simple non-mean-field model with trivial
statics and interesting dynamics, namely the parallel Monte Carlo 
dynamics of independent spherical spins in a quenched random field
~\cite{Nhammer}. We expect that both models lie in the same class as the 
lattice-gas models with kinetic constraints
of Kurchan, Peliti, and Sellitto
~\cite{KPS}~\cite{Sellitto}. The latter model can, however,
not be solved analytically.

In this work we shall give details underlying the
picture proposed in reference~\cite{Nhammer}.
In section \ref{General} we shall recall this picture.
In section \ref{Eqs} we shall derive dynamical equations for
averages, correlations and responses  
in a model of uncoupled harmonic oscillators
subject to Monte Carlo dynamics, introduced in reference~\cite{BPR}. 
In section \ref{OscSolution} we analyze these equations in the
non-equilibrium low temperature regime.
In section \ref{SpherSpins} we analyze the closely related model of
uncoupled spherical spins, introduced in ~\cite{Nhammer}.
We close with a discussion and summary.

\section{Thermodynamic picture for a system described by an 
effective temperature} \label{General}
\setcounter{equation}{0}\setcounter{figure}{0} 
\renewcommand{\thesection}{\arabic{section}.}

A state that slowly relaxes to equilibrium is characterized by 
$t$, the elapsed time, sometimes called ``age'' or ``waiting time''.
For glassy systems this is of special relevance.
For experiments on spin glasses it is known that non-trivial cooling
or heating trajectories can be described by an effective age~\cite{Hammann}.
Yet we do not wish to discuss spin glasses in this work. They have an
infinity of long time scales, or infinite order replica symmetry
breaking. Their phase transition is continuous, and involves 
power laws.

We shall restrict our treatment to systems with one diverging time 
scale, having, in the mean field limit, one step of replica symmetry
breaking. They are systems with first-order-type phase transitions,
with discontinuous order parameter, though usually there is no latent heat.
However, the same approach applies to true first order glassy
transitions that do have a latent heat. This occurs, for instance in the
transition from low density amorphous ice to high density amorphous
ice ~\cite{LDAHDAice}\cite{HES}. 
Theoretically such behavior occurs in spin
glasses in a transverse field, see e.g. 
~\cite{Mottishaw}\cite{Goldschmidt}
\cite{ThirumDobrov}\cite{NieuwRitort}.

We shall consider glassy transitions for liquids as well as for
random  magnets.
The results map onto each other by interchanging  volume $V$,
 pressure $p$, compressibility $\kappa=-\p \ln V/\p p$, and
expansivity $\alpha=\p \ln V/\p T$,
 by  magnetization $M$,  field $H$, susceptibility
$\chi=(1/N)\p M/\p H$,  and ``magnetizability'' $\alpha=(-1/N)\p M/\p T$,
respectively.

The picture to be investigated in this work starts
by describing a non-equilibrium state characterized by three parameters,
namely $T,H$ and the age {\it effective temperature} $T_e(t)$. 
As we shall see below, $T_e(t)$ enters naturally in the dynamical
solution of the problem.
For a set of smoothly related cooling experiments $T_i(t)$ 
at different fields $H_i$, one may express the effective
temperature as a continuous function: $T_{e,i}(t)$ $\to$ $T_e(T,H)$.
This sets a surface in $(T,T_e,H)$ space, that becomes multi-valued
if one first cools, and then heats. For covering the whole space one
needs to do very many experiments, e.g., at different fields
and at different cooling rates.
The results should agree with findings from heating experiments 
and aging experiments.  
Thermodynamics amounts to giving differential relations between 
observables at nearby points in this space.

For thermodynamics of glassy systems in the absence of currents,
all previous results can be summarized by expressing the change in heat
 as~\cite{NEhren} \cite{Nthermo}
\BEQ \label{dQ=}
\dbarrm Q=T\d\S+T_e\d\I
\EEQ
where $\S$ is the entropy of the fast or {\bf e}quilibrium 
{\bf p}rocesses  ($\beta$-processes)
and $\I$ the configurational entropy of the slow or ``configurational''
processes ($\alpha$-processes). This object is also known as
information entropy or complexity. Both $\S$ and $\I$ are state
functions in the sense that they depend on $T$, $T_e$, and on
$H$ or $p$. In particular, they are defined for any $T_e$, and,
within the present framework of one effective parameter, they do not
depend on the path along which this value was reached.
 
Notice that our separation 
in eq. (\ref{dQ=}) goes according to time scales. 
In the common use of the word, the configurational entropy $S_c$ 
is the entropy of the glass minus the entropy of the vibrational modes 
of the crystal~\cite{GibbsDiMarzio}.
For polymers, in particular, 
it still includes short-distance rearrangements, which
is a relatively fast mode. For the Gibbs-DiMarzio model 
it was confirmed numerically
that $S_c$ indeed does not vanish at any temperature,
thus violating the Adam-Gibbs relation 
$\tau\sim\exp({\rm const.}/S_c)$ between time scale and
configurational entropy~\cite{Binder}.
Our $\I$, on the other hand, only contains the slow components; 
the fast ones are supposed to be in equilibrium, and are counted in $\S$.
The properly formulated Adam-Gibbs relation should only refer to
slow quantities, so it should read: $\tau\sim \exp({\rm const}/\I)$.
Its applicability remains an open issue.
In a certain model glass with non-trivial fast and slow modes that
has a Kauzmann transition it is actually satisfied~\cite{NVFmodel}.

In the presence of currents eq. (\ref{dQ=}) would become 
$\dbarrm Q\le T\d\S+T_e\d\I$.
This decomposition is based on a system consisting of two
parts, with a slow exchange of heat between them, so having two time
 scales. A well known case is a cup of coffee at temperature $T_e$
in a room at temperature $T$. In that case $\I$ is the entropy of the
cup and  the coffee, 
$\S$ the entropy of the air and matter in the room, and $Q$ the 
heat of the combined system. To mention one case, 
cooling of the coffee in an isolated room will be described 
by $\dbarrm Q=0$ and  $T_e\d\I=-T\d \S<0$. 

It is both surprising and
satisfactory that a glass can be described by the same general law.
If also an effective pressure or field would be needed, then $\dbarrm Q$ 
is expected to keep the same form, but $\dbarrm W$ would change from its
standard value $-p\d V$ for liquids, or $-M\d H$ for magnets.
In the latter case it would become $-M_1\d H-M_2\d H_e$, where $H_e$
is the effective field, and $M_1$ and $M_2$ add up to $M$.
Such a complication could be needed in a larger class of systems.
It would make the picture technically a bit more difficult, and is
the subject of current research.

\subsection{First and second law}

For a glass forming liquid the first law $\d U=\dbarrm Q+\dbarrm W$
becomes
\BEA \label{thermoglassp}
\label{dUp=}\d U=T\d \S+T_e \d \I-p\d V
\EEA
One can define the free enthalpy
\BEA \label{Gp=} G&=&U-T\S-T_e \I+pV\EEA
that satisfies
\BEA
\label{dGp=}\d G&=&-\S\d T-\I\d T_e+V\d p
\EEA

The total entropy is 
\BEQ \label{Stot=}S=\S+\I \EEQ
(We should stress that the total entropy is not equal to $\S+T_e\I/T$;
there are many reasons why this unsymmetric form is incorrect. 
Let us mention that if the probability distribution decomposes into 
fast and slow processes as 
$P({\rm fast,\,slow})=P({\rm fast|slow})P({\rm slow})$, 
then the standard expression $S=-\tr P \ln P$ leads to (\ref{Stot=})
with $\S=\tr_{\rm slow}P({\rm slow})[-\tr_{\rm fast} 
P({\rm fast|slow})\ln P({\rm fast|slow})]$, just the 
entropy of the fast processes, averaged over the slow ones, 
and $\I=-\tr_{\rm slow}P({\rm slow})\ln P({\rm slow})$, just
the entropy of the slow processes.) 

The second law requires $\dbarrm Q\le T\d S$, so 
\BEQ (T_e-T)\d\I\le 0 \EEQ  

Since $T_e=T_e(T,p)$, and both entropies are functions of $T$, $T_e$ 
and $p$, the expression (\ref{dQ=}) yields a specific heat
\BEA
C_p&=&\frac{\p (U+pV)}{\p T}\fix_p=
T(\frac{\p \S}{\p T}\fix_{T_e,p}+
\frac{\p \S}{\p T_e}\fix_{T,p}\frac{\p T_e}{\p T}\fix_p)
+T_e(\frac{\p \I}{\p T}\fix_{T_e,p}+
\frac{\p \I}{\p T_e}\fix_{T,p}\frac{\p T_e}{\p T}\fix_p)
\EEA
In the glass transition region it holds that $T_e\approx T$.
Since the derivatives of $\S$ and $\I$ are smooth functions,
all factors, except $\p_T T_e$, are basically constant. This leads to
\BEA \label{CpTool}
C_p&=&C_1+C_2\frac{\p T_e}{\p T}\fix_p
\EEA
Precisely this form has been assumed half a century ago
by Tool~\cite{Tool}  as starting point for 
the study of caloric behavior in the glass formation region,
and has often been used for the explanation of experiments
~\cite{DaviesJones}\cite{Inamina}.  
It is thus a direct consequence of eq. (\ref{dQ=}).
Let us mention that Tool uses the term ``fictive temperature'' 
for $T_e$.

For magnetic systems the first law brings 
\BEA 
\label{dU=}\d U&=&T\d \S+T_e \d \I-M\d H
\EEA
One can define the free energy
\BEA \label{F=} F&=&U-T\S-T_e \I\EEA
that satisfies
\BEA\label{dF=}
\d F&=&-\S\d T-\I\d T_e-M\d H
\EEA

\subsection{Modified Maxwell relation}

For a smooth sequence of cooling procedures of a
glassy liquid, eq. (\ref{dUp=}) implies a modified
Maxwell relation between macroscopic observables
such as $U(t,p)\to U(T,p)= U(T,T_e(T,p),p)$ and $V$.
This solely occurs since $T_e$ is a non-trivial function of $T,p$ for
the smooth set of experiments under consideration. 

The consistency relation $\p^2 G/\p T\p p=\p^2 G/\p p\p T$
yields
\BEQ \label{consis}
-\frac{\p \S}{\p p}\fix_T -\frac{\p \I}{\p p}\fix_T\,
\frac{\p T_e}{\p T}\fix_p
=\frac{\p V}{\p T}\fix_p-\frac{\p \I}{\p T}\fix_p
\frac{\p T_e}{\p p}\fix_T\EEQ
Notice that difference relations as eq. (\ref{dU=}),
and the Legendre transformation that leads to (\ref{dF=}), 
do not invoke the functional dependence $T_e(T,p)$, since they
hold for any functional dependence, and even in absence of it. 
However, it does become relevant when dividing these equations 
by $\d T$ or $\d p$, as was done to derive (\ref{consis}).

Eq. (\ref{dUp=}) implies 
\BEQ T \frac{\p \S}{\p p}\fix_T=\frac{\p U}{\p p}\fix_T
-T_e\frac{ \p \I}{ \p p}\fix_T + p \frac{\p V}{\p p}\fix_T \EEQ
Eliminating $\p \S/\p p$ leads to
\BEQ \label{modMaxp}
\frac{\p U}{\p p}\fix_T + p\frac{ \p V}{\p p}\fix_T
+T\frac{\p V}{\p T}\fix_p
=T\frac{\p \I}{\p T}\fix_p\,\frac{\p T_e}{\p p}\fix_T-
T\frac{\p \I}{\p p}\fix_T\,\frac{\p T_e}{\p T}\fix_p+
T_e\frac{\p \I}{ \p p}\fix_T
\EEQ
This is the modified Maxwell relation between observables $U$ and $V$.
In equilibrium $T_e=T$, so the right hand side vanishes, and the
standard form is recovered.

Similarly, one finds for a glassy magnet
\BEQ \label{modMaxH}
\frac{\partial U}{\partial H}\fix_T+M-
T\frac{\partial M}{\partial T}\fix_H=
T_e\frac{\partial \I}{\partial H}\fix_T
+T\left(\frac{\partial T_e}{\partial H}\fix_T
\frac{\partial \I}{\partial T}\fix_H
-\frac{\partial T_e}{\partial T}\fix_H
\frac{\partial \I}{\partial H}\fix_T\right)
\EEQ

\subsection{Modified Clausius-Clapeyron relation}

Let us consider a first order transition between 
two glassy phases A and B. An example could be the transition from
low-density-amorphous ice to high-density-amorphous ice~\cite{LDAHDAice}.
For the standard Clausius-Clapeyron relation 
one uses that the free enthalpy $G$ is continuous along the
first order phase transition line $p_g(T)$. Since 
$T_e\neq T$, it is actually not obvious that $G$ 
should still be continuous there. The so far always confirmed
fact that in mean field models replica theory brings the
relevant physical free energy, leads us to expect that the
generalized free enthalpy (\ref{Gp=}) is indeed continuous.

Let us consider a first order transition between phases A and B,
that have their own $T_e$, $\S$ and $\I$. Let us denote the 
discontinuities in observables $O$ of the two states as
\BEQ \Delta O(T,p_g(T)) \equiv O_A- O_B \EEQ
Taking $O=G$ and differentiating $\Delta G=0$  one gets
\BEQ
\left(\Delta V-\Delta\left(\frac{\p T_e}{ \p p}\fix_T\I\right)
\right)\frac{\d p_g}{\d T}
=\Delta\S+\Delta \left(\frac{ \p T_e}{ \p T}\fix_p\I\right) 
\EEQ
$\S$ can be eliminated by means of (\ref{Gp=}).
Using again that  $\Delta G =0$, this yields
\BEQ\label{dUab=}
\Delta V \frac{\d p_g}{\d T} = \frac{\Delta U+p_g \Delta V}{T} +
\Delta\left(\frac{\d T_e}{ \d T}\I-\frac{T_e}{T}\I\right)
\EEQ
where $\d /\d T=\p/\p T+ (\d p_c/\d T)\p /\p p$ is the ``total'' 
derivative, i.e., the derivative along the transition line.
This is the modified Clausius-Clapeyron relation.
It would be very interesting to test this relation for ice.
For that substance
Mishima and Stanley~\cite{HES} have presented a thermodynamic
construction of the free enthalpy or Gibbs potential $G$.
It is, however, based on equilibrium ideas
and does not involve the effective temperature in the amorphous
phases. In particular, it assumes the validity of the original
Clausius-Clapeyron relation.
We feel that the results are not the physically relevant ones,
and that the analysis should be redone within our 
non-equilibrium thermodynamic framework.

When phase A is an equilibrium undercooled liquid, and phase B
is a glass, it holds that $T_e=T$ in phase A, and its $\I$-terms
 will cancel from (\ref{dUab=}), so this relation reduces to
\BEQ
\Delta V \frac{\d p_g}{\d T} = \frac{\Delta U+p \Delta V}{T} +
(\frac{T_e}{T}-\frac{\d T_e}{ \d T})\I
\EEQ
where $T_e$ and $\I$ are properties  of the glassy phase B.
Notice that eq. (7) of ref. ~\cite{NEhren} contains a misprint
in the prefactor of $\I$.

For standard glass forming liquids, there are no discontinuities
in $U$ and $V$.
It then holds that along the glass transition line $T_e(T,p_g(T))=T$,
implying $\d T_e/\d T=1$, which indeed removes  the $\I$ terms
from the last two relations.

\subsection{Ehrenfest relations and Prigogine-Defay ratio}

In the glass transition region a glass forming liquid exhibits 
smeared jumps in the specific heat $C_p$, the expansivity $\alpha$ 
and the compressibility $\kappa$. If one forgets about the smearing,
one may consider them as true discontinuities, yielding an analogy
with continuous phase transitions of the classical type.

Following Ehrenfest one may take the derivative of $\Delta
V(T,p_g(T))=0$. Using the definitions of $\alpha$ and $\kappa$, 
given  above, the result for a glass forming liquid may be written as
\BEQ \label{Ehren1p}
\Delta \alpha=\Delta \kappa \frac{\d p_g}{\d T}\EEQ
while for a glassy magnet
\BEQ \label{Ehren1H}
\Delta \alpha=\Delta \chi \frac{\d H_g}{\d T}\EEQ
The conclusion drawn from half a century of research on glass
forming liquids is that this relation is never satisfied
~\cite{DaviesJones}\cite{Goldstein}\cite{Jaeckle}~\cite{Angell}. 
This has very much hindered progress on a thermodynamical approach. 
However, from a theoretical viewpoint it is hard to imagine that
something could go wrong when just taking a derivative. 
We have pointed out that this relation is indeed satisfied
automatically~\cite{NEhren}, but it is important say what is
meant by $\kappa$ in the glassy state.

Let us make an analogy with spin glasses. In mean field theory
they have infinite order replica symmetry breaking. 
From the early measurements of Canella and Mydosh ~\cite{Mydoshboek} 
on AuFe it is known that 
the susceptibility depends logarithmically on the frequency, so on
the time scale. The short-time value, called Zero-Field-Cooled (ZFC)
susceptibility is a lower bound, while the long time value, called
Field-Cooled (FC) susceptibility is an upper bound. Let us
use the term ``glassy magnets'' for  spin glasses
with one step of replica symmetry breaking. They are relevant for comparison
with glass forming liquids. For them the situation is worse, as the 
ZFC value is discontinuous immediately below $T_g$.
( At $H=0$ one has $\chi_{ZFC}=\beta(1-q_{EA})$, while $\chi_{FC}=
\beta(1-(1-x_1)q_{EA})$ matches $\chi_{PM}=\beta$ at $x_1=1$.)
This occurs since giving the system more time to react on the field,
will lead to a really larger response, and it explains why already 
directly below the glass transition 
different measurements yield different values for $\kappa$.
These notions are displayed in figure 2.1

\begin{figure}[htb]
\label{chiplot}
\epsfxsize=10cm
\centerline{\epsffile{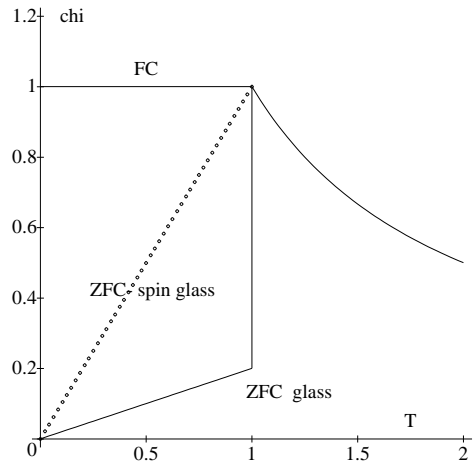}}
\caption{Schematic plot of the field-cooled (FC) and 
zero-field-cooled (ZFC) susceptibility in 
realistic spin glasses and  in glassy magnets, 
as function of temperature, in arbitrary units.
In realistic spin glasses the infinite time or field-cooled 
susceptibility is larger than the short time or zero-field-cooled
susceptibility. In magnetic analogs of realistic  glasses 
the short time susceptibility even has a smeared  discontinuity at 
the glass transition. In glass forming
liquids the same happens for the compressibility. }
\end{figure}

 Previous claims about the violation of the first Ehrenfest relation
can be traced back to the equilibrium thermodynamical idea that there
 is one, ideal $\kappa$,  to be inserted in (\ref{Ehren1p}).
Indeed, investigators usually considered cooling curves $V(T,p_i)$ 
at a set of pressures $p_i$ to determine $\Delta\alpha$ and 
$\d p_g/\d T$. (An alternative route, often followed in polymer physics, 
and leading to a very similar problem, is to change $p$ at 
many  constant values of $T$; then $\kappa$ depends strongly on the
rate of change of $p$.)
However, $\Delta \kappa$ was always determined in 
another way, often from measurements of the speed of sound,
or by making more complicated pressure steps~\cite{RehageOels}.
In equilibrium such alternative determinations would yield the
same outcome. In glasses this is not the case: the speed of sound is
a short-time process, and additional pressure steps modify the glassy
state.  Therefore alternative procedures are not allowed, and only
the cooling curves  $V(T,p_i)$ should be used. They constitute 
a liquid surface $V_{\rm liquid}(T,p)$ and a glass surface 
$V_{\rm glass}(T,p)$ in $(T,p,V)$ space. These surfaces intersect,
and the first Ehrenfest relation is no more than a mathematical
identity about the intersection line of these surfaces.
It is therefore automatically satisfied~\cite{NEhren}.
The most careful data we came across were collected by Rehage and 
Oels for atactic polystyrene~\cite{RehageOels}. 
In figure 2.2 
we present those data in a 3-d plot, underlining our point of view.

After submitting the original version of this paper, we realized that
McKenna has stressed that in experiments on glasses the 
isothermal compressibility differs from the isochoral compressibility
~\cite{McKenna}. He also concludes that alternative experiments are
not allowed, and that the first Ehrenfest relation indeed 
is merely a tautology.

\begin{figure}[htb]
\label{RehOelsplot3d}
\epsfxsize=15cm
\centerline{\epsffile{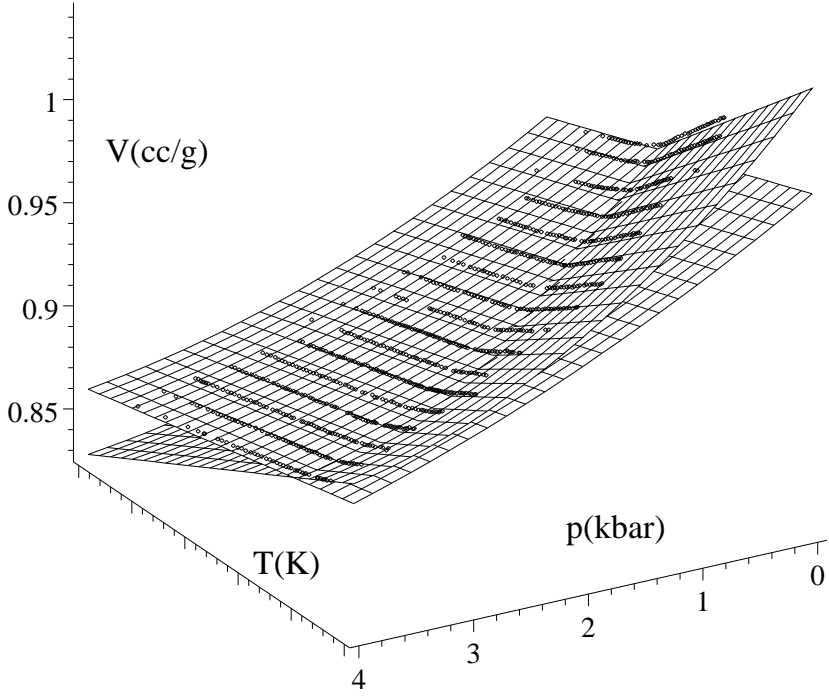}}
\vspace{-7cm}
\caption {Data of the glass transition for cooling
atactic polystyrene at rate 18 $K/h$, scanned 
from the paper of Rehage and Oels (1976):
specific volume $V$ ($cm^3/g$) versus temperature $T$
($K$) at various pressures $p$ ($k\,bar$). As confirmed by a polynomial
fit, the data in the liquid
essentially lie on a smooth surface, and so do the data in the glass.
The first Ehrenfest relation describes no more than the
intersection of these surfaces, and is therefore
automatically satisfied. The values for the compressibility
derived in this manner will generally differ from results obtained
via other procedures.}

\end{figure}

The second Ehrenfest relation derives from differentiating $\Delta
U(T,p_g(T))=0$. The obtained relation will also be satisfied automatically.
However, one then eliminates $\partial U/\partial p$ by means of
the Maxwell relation. In equilibrium this would yield
\BEQ \label{Ehren2pure}
\frac{\Delta C_p}{T_gV}=\Delta\alpha\frac{\d p_g}{\d T}
\EEQ
 We have already discussed that 
outside equilibrium it is modified, see eq. (\ref{modMaxp}). 
We thus obtain instead
\BEQ \label{modEhren2p} 
\frac{\Delta C_p}{T_gV}
=\Delta\alpha\frac{\d p_g}{\d T}
+\frac{1}{V}\left(1-\frac{\partial T_e}{\partial T}\fix_p\right)
\,\frac{\d \I}{\d T}
=\Delta\alpha\frac{\d p_g}{\d T}
+\frac{1}{V} \left(1-\frac{\partial T_e}{\partial T}\fix_p\right)
\left(\frac{\partial \I}{\partial T}\fix_p
+\frac{\d p_g}{\d T}\,\frac{\partial \I}{\partial p}\fix_T\right)
\EEQ
where $\d \I/\d T$
is the ``total'' derivative of the configurational entropy along the glass 
transition line. 
The last term  is new and vanishes only at equilibrium.
For magnets one gets
\BEQ \label{modEhren2H}
\frac{\Delta C}{NT}=\Delta \alpha\frac{\d H_g}{\d T}+
\frac{1}{N}\left(1-\frac{\partial T_e}{\partial T}\fix_H\right)
\left(\frac{\partial \I}{\partial T}\fix_H
+\frac{\d H_g}{\d T}\,\frac{\partial \I}{\partial H}\fix_T\right)
\EEQ
Along the glassy transition line the  equality 
$T_e(T,H_g(T))=T$ implies 
\BEQ \label{dTedT=1}
\frac{\d T_e}{d T}=\frac{\p T_e}{\p T}\fix_H+
\frac{\p T_e}{\p H}\fix_T\frac{\d H_g}{\d T}=1 \EEQ 

Combining the two original Ehrenfest relations one may eliminate the
slope of the transition line. This leads to consider the so-called
Prigogine-Defay ratio 
\BEQ\label{Pi=}
\Pi=\frac{\Delta C_p\Delta\kappa}{TV(\Delta \alpha)^2}
\EEQ
For equilibrium transitions it should be equal to unity. 
Assuming that at the glass transition a number of
unspecified parameters undergo a phase transition, Davies and Jones
showed that $\Pi\ge 1$~\cite{DaviesJones}, while DiMarzio showed 
that in that case the correct value is $\Pi=1$ ~\cite{DiMarzio}.
In glasses typical experimental values are reported in the range 
$2<\Pi<5$. It was therefore generally expected that $\Pi\ge 1$ is
a strict inequality.

We have pointed out, however, that, as the first Ehrenfest relation is
satisfied but the second is not, it holds that
\BEQ\label{Pip=}
\Pi=\frac{\Delta
C_p}{T V\Delta\alpha (\d p_g/\d T)}=1+
\frac{1}{V\Delta \alpha }
\left(1-\frac{\partial T_e}{\partial T}\fix_p\right) 
\frac {\d \I}{\d p} \EEQ
Depending on the set of experiments to be chosen, 
$\d p_g/\d T$ can be small or large, and $\Pi$ can also be below
unity. Rehage and Oels found $\Pi=1.09\approx 1$ at $p=1$
$k\,bar$, using a short-time value for $\kappa$~\cite{RehageOels}. 
Reanalyzing their data we find from (\ref{Pip=}), where the physically
relevant $\kappa$ has been inserted, a value $\Pi=0.77$, 
which is, surprisingly enough, below unity~\cite{NEhren}.

The definition (\ref{Pi=}) of $\Pi$ looks like a combination
of equilibrium quantities. This is misleading, however, since 
$\kappa_{\rm glass}$ depends sensitively on how the experiment is done. 
We conclude that the commonly accepted inequality $\Pi\ge 1$ is based 
on equilibrium assumptions. Our theoretical arguments and the
Rehage-Oels data show that such idea's are incorrect.
In particular this rules out the Gibbs-DiMarzio model as a 
principally correct model for the glassy state. 
It is an equilibrium model, and as such it will e.g. lead to $\Pi=1$,
in contradiction to experiments.

\subsection{Fluctuation formula}

The basic result of statistical physics is that it relates
fluctuations in macroscopic variables to response of their averages
to changes in external field or temperature.
We have wondered whether such relations generalize to the glassy
state. We have found arguments in favor of such a possibility
both from the fluctuation-dissipation relation and by
exactly solving the dynamics of model systems ~\cite{Nhammer}.
Susceptibilities appear to have a non-trivial decomposition, that
looks as being very general. Here we give  arguments leading to it.

Later we shall consider models where two fields $H_a$ ($a=1,2)$ 
are present, and two magnetizations $M_a$ ($a=1,2$).
In cooling experiments at fixed field $H=(H_1,H_2)$ it holds that 
$M_a=$$M_a(T(t),T_e(t,H),H)$. For thermodynamics one eliminates
time to express $T_e(t,H)\to T_e(T,H)$, implying
$M_a=$$M_a(T,T_e(T,H),H)$. One may then expect three terms:
\BEA\label{flucts=}
\chi_{ab}&\equiv&\frac{1}{N}\,
\frac{\partial M_a}{\partial H_b}\Bigl|_T\Bigr.
= \chi_{ab}^{\rm fluct}(t)+\chi_{ab}^{\rm loss}(t)+\chi_{ab}^{\rm conf}(t)
\EEA
The first two are defined by
\BEA \label{chifluct1}
\chi_{ab}^{\rm fluct}(t)+\chi_{ab}^{\rm loss}(t)
&=&\frac{1}{N} \frac{\partial M_a}{\partial H_b}
\Big|_{T,T_e}\Bigr. \EEA
To calculate them separately, we switch from a cooling experiment
to an aging experiment at the considered
$T$, $T_e$ and $H$, by keeping, in Gedanken, $T$ fixed from then on. 
The system  will continue to age,  expressed by $T_e=T_e(t;T,H)$. 
We may then use the equality
\BEQ\label{Mabhelp}
\frac{\partial M_a}{\partial H_b}\Big|_{T,t}\Bigr. 
=\frac{\partial M_a}{\partial H_b}
\Big|_{T,T_e}\Bigr.+ \frac{\partial M_a}{\partial T_e}
\Big|_{T,H}\Bigr. \frac{\partial T_e}{\partial H_b}
\Big|_{T,t}\Bigr. 
\EEQ
We have conjectured ~\cite{Nhammer}
that the left hand side may be written as the
sum of fluctuation terms for fast and slow processes,
\BEQ\label{Mabhelp1}\label{chifluct}
\chi_{ab}^{\rm fluct}(t)=
\frac{1}{N}\frac{\partial M_a}{\partial H_b}\Big|_{T,t}\Bigr. 
=\frac{ \langle \delta M_a(t)\delta M_b(t)\rangle_{\rm fast}}{NT(t)}+
\frac{\langle \delta M_a(t)\delta M_b(t)\rangle_{\rm slow}}{NT_e(t)}\EEQ
The first term is just the standard equilibrium expression for the fast
equilibrium processes.
Notice that slow processes enter with their own temperature,
the effective temperature.  This decomposition  
is confirmed by use of the fluctuation-dissipation relation 
in the form to be discussed below. 
Combination of (\ref{chifluct1}), (\ref{Mabhelp}) and
(\ref{Mabhelp1}) yields
\BEA \label{chiloss}
\chi_{ab}^{\rm loss}(t)&=&
-\frac{1}{N}\frac{\partial M_a}{\partial T_e}\Big|_{T,H}\,\,
\frac{\p T_e}{\partial H_b}\Bigr|_{T,t}
\EEA
The fluctuation terms are instantaneous, and thus the same for aging
and cooling. The loss term is a correction, 
related to an aging experiment. It measures the decrease of
fluctuations below the glass transition, which will be small in 
the models to be discussed later on.

In the models to be considered below, dynamics 
in the glassy phase is essentially independent of the actual $T$, 
leaving almost no difference between cooling and aging.
This is due to the simplicity of the model.

Since $T_e\neq T$, there occurs in eq. (\ref{flucts=})
also a new, configurational term
\BEA \label{chiconf}
\chi_{ab}^{\rm conf}=\frac{1}{N}\,
\frac{\partial M_a}{\partial T_e}\Bigl|_{T,H}\,\,
\frac{\partial T_e}{\partial H_b}\Bigr|_{T}
\EEA
It originates from the difference in  the system's
structure for cooling experiments at nearby fields.
For  glass forming liquids such a term occurs in the compressibility.
Its existence was anticipated in some earlier works.
Goldstein ~\cite{Goldstein} points out that
$V_{\rm glass}$ depends stronger on the pressure of formation 
$p_{\rm form}$ than on
the one remaining after partial release of pressure, $p_{\rm final}$.
J\"ackle ~\cite{Jaeckle} 
then assumes that for infinitely slow cooling $p_{\rm form}$
is the only additional system parameter, and argues that
$\Delta \kappa_T\to\Delta\kappa= \Delta \kappa_T
+\partial \ln V/\partial p_{\rm form} =\Delta\alpha\,\d T_{g}
/\d p$ and that this implies $\Pi=\Delta \kappa_T/\Delta \kappa>1$.
He thus also considers one extra system variable, and also 
argues the existence of a configurational term. 
We do not wish to restrict to adiabatically slow cooling,
and we do not agree with his conclusion on $\Pi$. 
Notice that our approach allows, in principle, to find 
the configurational term (\ref{chiconf}) for typical
cooling procedures from construction 
of $V(T,T_e,p)$ in full $(T,T_e,p)$-space.

From the analysis to be given below, we find no reason why such 
universal quasi-equilibrium relations could also hold 
between the specific heat and the energy fluctuations.
In the models of the present paper
the energy fluctuation are smaller by one order of magnitude, 
and model dependent.
The absence of such a general relation allowed us to apply the 
very same two-temperature 
approach to black holes, without obtaining a contradiction with
their negative specific heat~\cite{Nblackhole}.

\subsection{Fluctuation-dissipation relation}

Nowadays quite some attention is payed to the fluctuation-dissipation
relation in the aging regime of glassy systems. It was first put forward
in works by Horner~\cite{Horner1}\cite{Horner2}\cite{CHS} 
and then by Cugliandolo and Kurchan ~\cite{CuKu}. This relation has
become a central point in research on off-equilibrium systems,
for a review, see ~\cite{BCKM}.

Our formulation is that in the aging regime there holds the following
relation between the cross correlation 
$C_{ab}(t,t')=\langle O_a(t)O_b(t')\rangle-
\langle O_a(t)\rangle\langle O_b(t')\rangle$ of macroscopic observables
$O_a(t)$ and $O_b(t')$, and the response 
$G_{ab}(t,t')=\delta\langle O_a(t)\rangle/\delta H_b(t')$
of $O_a(t)$ to a short, small 
field change applied at an earlier time $t'$:
\BEQ \label{FDR=}
\frac{\partial C_{ab}(t,t')}{\p t'}
=\tilde T_e(t'){G_{ab}(t,t')}\EEQ
with $\tilde T_e(t)$ being the effective temperature for the
FDR, while in the equilibrium or
short-time regime, $T$ replaces $\tilde T_e$. This relation has
been confirmed numerically e.g. for a soft sphere glass~\cite{Parisi}.
It is remarkable that the $\tilde T_e$ 
is a function of one of the times only. However, one should keep
in mind that $C$ and $G$ typically  have a $t'/t$ scaling, while
$\tilde T_e$ typically is a smooth function of $\ln t$, 
a variable that basically equals $\ln t'$.

One expects that $\tilde T_e(t)$ is close to the 
``thermodynamic'' effective temperature $T_e(t)$.
Let us show how this comes about.

The two fluctuation terms in eq. (\ref{chifluct}) are consistent with
(\ref{FDR=}). To prove this, let us neglect switching effects
(see section 3.4), and use the definition 
\BEQ\label{Mabhelp14}
\frac{1}{N}\,\frac{\partial M_a}{\partial H_b}\Big|_{T,t}\Bigr. 
=\int_0^t G_{ab}(t,t')\d t' \EEQ
We split the integral up in the regions $(t-\tau_\beta,t)$
and $(0,t-\tau_\beta)$, where $\tau_\beta$ is the time after which 
the fast or $\beta$ processes have died out. Their contribution
has the equilibrium form, while in the second interval we
may insert (\ref{FDR=}), which yields
\BEQ\label{Mabhelp15}
\frac{1}{N}\,\frac{\partial M_a}{\partial H_b}\Big|_{T,t}\Bigr. 
=\frac{ \langle \delta M_a(t)\delta M_b(t)\rangle_{\rm fast}}{NT(t)}+
\int_0^{t-\tau_\beta} \d t'\frac{1}{\tilde T_e(t')} \,\frac{\p C_{ab}(t,t')}
{\p t'} \EEQ
We perform a partial integration, and can neglect
the value at the lower boundary $t=0$. In the remaining
term we insert a factor $1=\p_{t'}C_{ab}(t,t')/\p_{t'}C_{ab}(t,t')$.
We can then do another partial integral, and we could, in principle,
repeat this process. All terms at $t-\tau_\beta$ share a common
factor, namely the plateau value of $C_{ab}(t,t')$,
\BEQ C_{ab}^{\rm plateau}(t)\equiv C_{ab}(t,t-\tau_\beta)=
\frac{\langle \delta M_a(t)\delta M_b(t-\tau_\beta)\rangle_{\rm slow}}{N}
\EEQ
that also enters the relation
\BEQ
\frac{1}{N}\langle \delta M_a(t)\delta M_b(t)\rangle_{\rm fast}
=C_{ab}(t,t)-C_{ab}^{\rm plateau}(t)
\EEQ
As a result, we derive from (\ref{FDR=}) our Ansatz (\ref{chifluct})
with a factor 
\BEQ
\frac{1}{T_e(t)}=\frac{1}{\tilde T_e(t)}+ 
\frac{\p_t \tilde T_e(t)}{\tilde T_e^2(t)}\,\frac{C_{ab}(t,t')}
{\p_{t'} C_{ab}(t,t')}\fix_{t'=t-\tau_\beta}+\cdots \EEQ
This may be inverted, to yield
\BEQ \label{tildeTegen}
\tilde T_e(t)=T_e(t)+\dot T_e(t)\left(\frac{ \p \ln C_{ab}(t,t')}{\p t'}
\fix_{t'=t-\tau_\beta}\right)^{-1}+\cdots
\EEQ
It is clear that the effective temperatures $T_e$ and $\tilde T_e$
are not identical. However, in the models to be analyzed later on,
we shall find  that the difference is small.

Notice that the ratio $\p_{t'}C(t,t')/G(t,t')=\tilde T_e(t')$ 
is allowed to depend on time $t'$.
The situation with constant $T_e$ ($=T/x$, with $x$ the break point
of the Parisi function) is well known from mean
field spin glasses, but we shall not find such a
constant $T_e$ in the models to be studied.
In ~\cite{BCKM} it is
reviewed that in mean field spin glasses the fluctuation-dissipation
parameter $X(t,t')$$\equiv$$ TG(t,t')/\partial _{t'}C(t,t')$ simplifies to
$X(t,t')\equiv \hat X(C(t,t'),t')$$=\hat X(0,t')\to
const$. As our $T_e(t')$ will depend logarithmically on time,
the $t'$-dependence of our $\hat X(0,t')=T/T_e(t')$ cannot be neglected.
We can therefore conclude that such a time-independence is an artifact
of the mean field approximation. This supports our earlier conclusion
that only at exponential time scales $\sim\exp(N)$ the dynamics of the
mean field spin glass is related to that of realistic systems~\cite{Nthermo}.
In the numerical evaluation of the ``fluctuation-dissipation
ratio'' $T/T_e$  one should therefore keep in mind the realistic 
possibility of a slow time dependence of $T_e$.

\subsection{Time-scale arguments}

Consider a simple system that has only one type of processes ($\alpha$
processes), which falls out of equilibrium at some low $T$. 
When it ages a time $t$ at $T=0$ it will have achieved a state
with effective temperature $T_e$, that can be estimated
by equating time with the equilibrium timescale. Let us define 
${\overline  T}_e$ by
\BEQ \label{barTe}
t=\tau_{eq}({\overline T}_e)\EEQ
We shall check in the models to be studied below that, 
to leading order in $\ln t$, it holds that ${\overline T}_e=T_e$. 
(The first non-leading order turns out to be  non-universal, 
since it already depends on  numerical prefactors of $\tau_{eq}$.) 
This equality also
is found in cooling trajectories, when the system is well
inside the glassy regime. It says that the
system basically has forgotten its history, and ages on its own,
without caring about the actual temperature.
We feel that this is caused by the fact that dynamics in each new
time-decade is basically independent of previous decade.

In less trivial systems, for instance those having a
Vogel-Tammann-Fulcher law,  the timescale may have parameters
that depend on the actual temperature, implying $\tau_{eq}=\tau(T,T_e)$.
We then expect that, to leading order, $T_e$ follows by
equating this expression with time $t$. 

In many systems one finds a $t'/t$ scaling 
in the aging regime of two-time quantities. 
There is a handwaving argument to explain that:
\BEQ
C(t,t')\approx C(\frac{t-t'}{\tau_{eq}(T_e(t'))})
\approx C(\frac{t-t'}{t'})=C(\frac{t}{t'})\EEQ
showing indeed the familiar $t/t'$ scaling. In the models to be
studied below we shall find logarithmic  scaling corrections. They
become strong at low $T$, and 
change the $\sqrt{t'/t}$ decay at $T>0$ to a $t'/t$ decay at $T=0$. 
So this argument might  apply only to a subset of systems 
that fall within the scope of our approach.

\subsection{Results for simple systems that become glassy near $T=0$} 
In the remainder of this paper we shall consider two simple systems
 having only one type of processes ($\alpha$ processes), 
which fall out of equilibrium at some low $T$. 
Then the effective temperature $T_e(t)$ is expected to show up in 
the following deviations from  the equilibrium situation:
\begin {itemize}
\item matching the internal energy: $U(t,H)=U_{eq}(T_e(t),H)$
\item matching the magnetization $M(t,H)=M_{eq}(T_e(t),H)$
\item from the configurational entropy via $\d U=T_e\d \I-M\d H$
\item matching time with the equilibrium timescale:
  $t\sim \tau_{eq}(T_e(t))$
\item via the fluctuation formula $\chi^{\rm fluct}=\beta_e(t)  
\langle \delta M^2(t) \rangle$
\item from the fluctuation-dissipation relation:
$\frac{\partial C(t,t')}{\partial t'}=\tilde T_e(t')G(t,t')$
\end{itemize}
Even though these relations are not all independent, it is pretty
clear that the whole glassy dynamics is strongly governed by one
parameter: the effective temperature.

\section{Monte Carlo dynamics of uncoupled harmonic oscillators}
\label{Eqs}
\setcounter{equation}{0}\setcounter{figure}{0} 
\renewcommand{\thesection}{\arabic{section}.}

Bonilla, Padilla and Ritort have recently considered an exactly
solvable model with slow dynamics~\cite{BPR}. It showed interesting,
glassy behavior at low temperatures. In this section we present
many details and further results for this model. This will 
also be a pedogical step for the analysis of the spin model
of section 5.

After including an external field, the Hamiltonian reads
\BEQ
\H=\half K\sum_ix_i^2-H\sum_ix_i\equiv \half KM_2-HM_1
\EEQ
where
\BEQ \label{M1M2def}
M_k=Nm_k=\sum_ix_i^k
\qquad (k=1,2)
\EEQ
We 
introduce the shifted energy
\BEQ
E=N\e 
=\H+N\frac{H^2}{2K}
\EEQ
In a Monte Carlo step with parallel updates
one replaces $x_i\to x_i'=x_i+r_i/\sqrt{N}$, 
defining $M_k'=\sum(x_i+r_i/\sqrt{N})^k$.
The thermal noise variables $r_i$ are independent Gaussian
random variables with 
average zero and variance $\Delta^2$. For a parallel update of all $x_i$, 
this leads to the noise averaged transition probability
from a state with $(M_1,M_2)$ to states with
$(M_1',M_2')$ $=$ $(M_1+y_1,M_2+y_2)$ 
\BEA
P(y_1,y_2|M_1,M_2)&=&
\left(\prod_i \int_{-\infty}^\infty 
\frac{\d r_i}{\sqrt{2\pi\Delta^2}}e^{-\frac{r_i^2}{2\Delta^2}}\right)\,
\delta(M_1'-M_1-y_1)\delta(M_2'-M_2-y_2)\nn
&=&\frac{1}{4\pi \Delta^2\sqrt{m_2-m_1^2}}
\exp[-\frac{y_1^2}{2\Delta^2}-\frac{(y_2-\Delta^2-2y_1m_1)^2}
{8\Delta^2(m_2-m_1^2)}] \nn
&\equiv& p(y_1,y_2|m_1,m_2)
\EEA
where we took the convention that probabilities involving  
extensive parameters are written with capitals, while those 
that involve intensive parameters are written in lower case.
To derive this result, the delta-functions have been written in a
plane-wave representation, and the limit $N\to\infty$ has been taken.
Setting $y_1=y$ we introduce the variable $x$ by
\BEQ
x=\frac{K}{2}y_2-Hy_1;\qquad y_2=\frac{2}{K}(x+Hy)\EEQ
A Monte Carlo move implies a change $E'=E+x$. 
The transition probability  may be decomposed as
\BEA\label{Pxyem=}
P(y,y_2|M_1,M_2)\d y\d y_2&=&P(x,y|E,M_1)\d x\d y=
p(x|\e)p(y|x,\e,m_1)\d x \d y
\EEA
with conditional probabilities
\BEA \label{PxPy=}
p(x|\e)&=&  \frac{1}{\sqrt{2\pi\Delta_x }}\,
\exp({-\frac{(x-x_0)^2}{2\Delta_x}}) \\
p(y|x,\e,m_1)&=&\frac{1}{\sqrt{2\pi \Delta_y}}
\,\exp({-\frac{(y-y_0)^2}{2\Delta_y}})
\EEA
having parameters
\BEA
&x_0= \frac{\Delta^2 K}{2};\qquad &\Delta_x
=\Delta^2(K^2m_2-2HKm_1+H^2)=2K\Delta^2\e \EEA
\BEA
y_0=\frac{-(K\Delta^2-2x)(Km_1-H)}{2(K^2m_2-2HKm_1+H^2)}
=\frac{K\Delta^2-2x}{4\e}\mu_1 ;\qquad
\Delta_y=\frac{\Delta^2K^2(m_2-m_1^2)}{K^2m_2-2HKm_1+H^2}
=\Delta^2(1-\frac{K\mu_1^2}{2\e})
\EEA
where we defined the deviations from equilibrium
\BEQ\label{eps=mu1=}
 \varepsilon=\half K(m_2-m_1^2+\mu_1^2),\qquad\mu_1=\frac{H}{K}-m_1\EEQ
We shall frequently encounter the energy scale
\BEQ \label{A=}
 A=\frac{\Delta^2K}{8}
\EEQ

In one Monte Carlo step the probability of $E$ evolves as ~\cite{BPR}
\BEA \label{PEp}
P(E',t+\dt)&=&\int\d E P(E,t)\int \d x p(x|\frac{E}{N})
[W(\beta x)\delta(E'-E-x)+(1-W(\beta x))\delta(E'-E)] \nn
&=&P(E',t)+\int \d E P(E,t)\int \d x p(x|\frac{E}{N})
W(\beta x)[\delta(E'-E-x)-\delta(E'-E)]
\EEA
here $W(\beta x)=1$ for $x<0$ and $W(\beta x)=e^{-\beta x}$ for $x>0$ 
is the Metropolis 
acceptance rate. The second term describes the rejected moves.
Notice that the energy aspects are independent of the field $H$, 
for the physical reason that $H$ merely causes a shift of the 
equilibrium position, but not in $E=(K/2)\sum_i(x_i-H/K)^2$.
This is due to the simplicity of the model.

If  one also keeps track of the magnetization, one has
\BEA
P(E',M',t+\dt)=P(E',M',t)+
&\int& \d E\d MP(E,M,t)\int \d x\d y P(x,y|E,M)W(\beta x)\nn
&\times& [\delta(E'-E-x)\delta(M'-M-y)-\delta(E'-E)\delta(M'-M)]
\EEA
which, of course, is consistent with $P(E,t)=\int \d M P(E,M,t)$.

\subsection{Evolution of average observables}

We can now calculate evolution of physical observables. 
One derives from (\ref{PEp})
\BEA\label{dEdt}
\langle E(t+\dt)\rangle&=&\int \d E'\d M' E' P(E',M',t+\dt)=
\langle E(t)\rangle+\int \d E\d x W(\beta x)P(E,t)xp(x|\frac{E}{N})
\EEA
where $\langle E(t)\rangle$ arises from the term without $W(\beta x)$.
In the thermodynamic limit (i.e. for large $N$) 
 $P(E,t)$ will be sharply peaked around $\langle E(t)\rangle$, so one
obtains a closed equation for the scaled average
$\e(t)=\langle E(t)\rangle/N$~\cite{BPR}
\BEQ \label{dedt=}
\frac{\d\e(t)}{\d t}=\int \d x \,W(\beta x)x\, p(x|\e(t))
\EEQ
This simplifying property is due to the lack of interaction
between the oscillators.
 
In the same way we proceed for the evolution of the magnetization 
\BEA\label{dMdt}
\langle M(t+\dt)\rangle&=&
\langle M(t)\rangle+\int \d E\d M\d x\d yW(\beta x)
P(E,M,t)yp(x|\frac{E}{N})
p(y|x,\frac{E}{N},\frac{M}{N})
\EEA
Here, and everywhere in the sequel, the $y$-integrals are Gaussian,
and can be carried out analytically. This makes the problem with a
field hardly more complicated  than without. 
We obtain
\BEQ \label{dmdt=}
\frac{\d m(t)}{\d t}=\int \d x \,W(\beta x)\,y_0\,p(x|\e(t))
=-(m(t)-\frac{H}{K})f(t)\EEQ
where
\BEQ\label{f=}
f(t)=-\int_\minfty^\infty \d x W(\beta x)\frac{\p y_0}{\p m}
p(x|\e(t))
=\int_\minfty^\infty \d x W(\beta x)\frac{4A-x}{2\e(t)}p(x|\e(t))
\EEQ

\subsection{Fluctuations}

The evolution of bilinear forms is a bit more involved. Let us
consider the energy fluctuations. One has 
\BEQ \langle E^2(t+\dt)\rangle=\langle E^2(t)\rangle+
\int \d E\d x W(\beta x)P(E,t)(2xE+x^2)p(x|\frac{E}{N})
\EEQ
Using (\ref{dEdt}) this may be written as
\BEQ \langle E^2(t+\dt)\rangle-\langle E^2(t)\rangle
-2\langle E(t)\rangle(\langle E(t+\dt)\rangle-\langle E(t)\rangle)=
\int \d E\d x W(\beta x)P(E,t)p(x|\frac{E}{N})(2x\delta E+x^2)\EEQ
where $\delta E=E-\langle E(t)\rangle$.
Expanding $p(x|E/N)$ around $E=\langle E(t)\rangle$ one obtains 
for large $N$
\BEQ\label{dDE2dt=}
\frac{\d}{\d t}
\frac{ \langle\delta E^2\rangle}{N}
=\int \d x W(\beta x) 
(\frac{\langle\delta E^2\rangle}{N}2x\frac{\partial}{\partial \e}
+x^2)p(x|\e)
\EEQ
In the same way one derives for the evolution equation for
fluctuations in $M$
\BEA 
\frac{\d}{\d t}\,
\frac{ \langle\delta M^2\rangle}{N}
&=&\int \d x \d y W(\beta x)
(\frac{\langle\delta M^2\rangle}{N}2y\frac{\partial}{\partial m}
+\frac{\langle\delta E\delta M\rangle}{N}2y\frac{\partial}{\partial \e}
+y^2) p(x|\e)p(y|x,\e,m)\EEA
The $y$-integral is Gaussian, and can be carried out.
The result reads
\BEA \label{MMeq}
\frac{\d}{\d t}\,\frac{ \langle\delta M^2\rangle}{N}&=&\int \d x
W(\beta x) 
(\frac{\langle\delta M^2\rangle}{N}\frac{\partial}{\partial m}2y_0
+\frac{\langle\delta E\delta M\rangle}{N}\frac{\partial}{\partial\e}2y_0
+y_0^2+\Delta_y) p(x|\e)
\EEA
while for the cross-correlations
\BEA \label{EMcorr}
\frac{\d}{\d t}\,
\frac{ \langle \delta E\delta M\rangle}{N}
&=&\int \d x \d y W(\beta x) (
\frac{\langle\delta M^2\rangle}{N}x\frac{\partial}{\partial m}+
\frac{\langle\delta E\delta M\rangle}{N}
(x\frac{\partial}{\partial \e}+
y\frac{\partial}{\partial m})
+\frac{\langle \delta E^2\rangle}{N}y\frac{\partial}{\partial \e}
+xy)p(x|\e)p(y|x,\e,m) \nn &=&
\int \d x W(\beta x) (
\frac{\langle\delta E\delta M\rangle}{N}
(x\frac{\partial}{\partial \e}+
\frac{\partial}{\partial m}y_0)
+\frac{\langle \delta E^2\rangle}{N}\frac{\partial}{\partial \e}y_0
+xy_0)p(x|\e)
\EEA
Recalling that $M=M_1$ and the definition (\ref{M1M2def})
of $M_2$, and adopting the definition of correlators to be given in
eq. (\ref{mambdef}), we may also cast these results in the form 
\BEA
\frac{\d}{\d t}C_{ab}(t,t)=\int \d x W(\beta x)\left\{ \right.
{ {\y_a}}\,\,{{\y_b}}
+\Delta_y\left(-\frac{H_1}{H_2}\right)^{a+b-2}
+ \sum_{c=1}^2\frac{\partial}{\partial m_c}\left({\y_a}C_{cb}(t,t)
+{\y_b}C_{ca}(t,t)\right) \left.\right\}p(x|m_1,m_2) 
\EEA
where $a,b$ $=$ $1,2$, $H_1=H$, $H_2=-K/2$,
and, in the present model, $p(x|m_1,m_2)=p(x|\e)$,
with $\e$ defined in (\ref{eps=mu1=}). Furthermore, 
\BEQ {{\y_1}}=y_0\qquad { {\y_2}}=-\frac{x+H_1y_0}{H_2}\EEQ
Since $\delta E=(K/2)\delta M_2-H\delta M_1$$=$$-\sum_cH_c\delta M_c$, 
previous results are recovered from these relations.

\subsection{Correlation and response functions}
One can also consider the evolution of two-time quantities. 
The correlation and response functions for magnetization and energy
are defined as 
\BEA \label{mmdef}
&C_{ m m}(t,t')=\frac{1}{N}\,\langle \delta M(t)\delta M(t')\rangle 
\qquad\qquad
&G_{ m m}(t,t')=\frac{1}{N}\,\frac{\delta\langle M(t)\rangle}
{\delta H(t')} \\ \label{emdef}
&C_{\e m}(t,t')=\frac{1}{N}\langle \delta E(t)\delta M(t')\rangle 
\qquad\qquad
&G_{\e m}(t,t')=\frac{1}{N}\,\frac{\delta\langle E(t)\rangle}{\delta H(t')} \\
\label{medef}
&C_{ m\e}(t,t')=\frac{1}{N}\langle \delta M(t)\delta E(t')\rangle 
\qquad\qquad
&G_{ m\e}(t,t')=\frac{T(t')}{N}
\frac{\delta\langle M(t)\rangle}{\delta T(t')}\\ 
\label{eedef}
&C_{\e\e}(t,t')=\frac{1}{N}\langle \delta E(t)\delta E(t')\rangle 
\qquad\qquad
&G_{\e\e}(t,t')=\frac{T(t')}{N}\,
\frac{\delta\langle E(t)\rangle}{\delta T(t')} 
\EEA
In eq. (\ref{M1M2def}) we have introduced the 
macroscopic observables $M_1=M$ and $M_2$.
The related correlators and responses are
\BEA \label{mambdef}
&C_{ab}(t,t')=\frac{1}{N}\,\langle \delta M_a(t)\delta M_b(t')\rangle 
\qquad\qquad
&G_{ab}(t,t')=\frac{1}{N}\,\frac{\delta\langle M_a(t)\rangle}
{\delta H_b(t')}\qquad (a,b=1,2) \EEA
They code the same information, but will be more useful at some stages.
$C_{mm}$ is just another notation for $C_{11}$, 
while $C_{\e m}=(K/2)C_{21}-HC_{11}$ and
$C_{\e\e}=(K^2/4)C_{22}-(HK/2)(C_{12}+C_{21})+H^2C_{11}$. 
Similar relations hold for the $G$'s:
$G_{\e m}=(K/2)G_{21}-HG_{11}$, $G_{m\e}=-(K/2)G_{12}-HG_{11}$,
and $G_{\e\e}=(4/K^2)G_{22}+(HK/2)G_{12}-(HK/2)G_{12}+H^2G_{11}$, where we
used that $G_{12}(t,t')=-G_{21}(t,t')$.

To derive the evolution of the correlations one considers
\BEA \langle M(t+\dt)M(t')\rangle 
&\equiv& \int \d E'\d M' \d E_1\d M_1\,P(E',M',t+\dt;E_1,M_1,t')M'M_1 \\
&=&\langle M(t)M(t')\rangle + \int \d E\d M \d E_1\d M_1 \dx\dy
W(\beta x) P(E,M,t;E_1,M_1,t') P(x,y|E,M)yM_1 \nonumber
\EEA
Subtracting $\langle M(t+\dt)-M(t)\rangle\langle M(t')\rangle$
and expanding $P(x,y|E,M)$,
this yields the evolution equation for $C_{mm}$. The $y$-integral
can again be performed. In similar ways one proceeds for
the other correlation functions. One finally has:
\BEA\label{dCmmdt=}
\frac{\partial}{\partial t}\,C_{mm}(t,t')
&=&\int \dx W(\beta x)
[C_{mm}(t,t')\frac{\partial}{\partial m}+
C_{\e m}(t,t')\frac{\partial}{\partial \e}]y_0p(x|\e)
\EEA
\BEA \label{dCemdt=}
\frac{\partial}{\partial t}\,C_{\e m}(t,t')
&=& \int \dx  W(\beta x)x
C_{\e m}(t,t')\frac{\partial}{\partial \e}p(x|\e)
\EEA
\BEA\label{dCmedt=}
\frac{\partial}{\partial t}\,C_{m\e}(t,t')
&=&\int \dx W(\beta x)
[C_{m\e}(t,t')\frac{\partial}{\partial m}+
C_{\e\e}(t,t')\frac{\partial}{\partial \e}]y_0p(x|\e)
\EEA
\BEA\label{dCeedt=}
\frac{\partial}{\partial t}\,C_{\e\e}(t,t')=\int \dx  W(\beta x)x
\frac{\partial}{\partial \e}p(x|\e)\,C_{\e\e}(t,t')
\EEA
Their equal-time values follow from the above fluctuation formulae. 

The equivalent formulation is 
\BEA 
\frac{\p}{\p t}C_{ab}(t,t')=
\sum_c C_{cb}(t,t')\frac{\partial}{\partial m_c}
 \int\d x W(\beta x) \overline {y}_a p(x|m_1,m_2)
\EEA

\subsection{Response functions}

The energy-energy response function $G_{\e\e}(t,t')$, 
defined in (\ref{eedef}), takes the form
\BEA G_{\e\e}(t,t')=\frac{-\beta}{N}\int \d E \d E_1 \dx_1 E 
[P(E,t|E_1+x_1,t'+\dt)-P(E,t|E_1,t'+\dt)]
\frac{\partial W(\beta x_1)}{\d t\,\,\partial \beta}
p(x_1|\frac{E_1}{N})P(E_1,t')
\EEA
For our parallel Monte Carlo updates it holds that $\d t=1/N$.
Both terms satisfy the same evolution equation, implying 
\BEA\label{dGeedt=}
\frac{\partial}{\partial t}\,G_{\e\e}(t,t')=-g(t)\,G_{\e\e}(t,t')\EEA
with
\BEA
g(t)=-\int \dx  W(\beta x)x\frac{\partial }{\partial \e}  p(x|\e)
\EEA
Since in the oscillator model the energy evolves independently of $H$, 
it is obvious that $G_{\e m}=0$ at all times. 
The relation $G_{\e m}=-\sum_c H_c G_{c\,1}$
then implies $G_{21}(t,t')=-(H_1/H_2)G_{11}(t,t')$. 
In the spherical spin model, to be introduced later, this 
argument does not hold.

From the evolution (\ref{dmdt=}) of $m(t)$ one gets immediately that
$G_{mm}$ and $G_{m\e}$ satisfy
\BEA \label{Gmmttp=}
 \frac{\partial}{\partial t}\,G_{mm}(t,t')&=&-f(t)G_{mm}(t,t');
\qquad  \frac{\partial}{\partial t}\,G_{m\e}(t,t')=-f(t)G_{m\e}(t,t')
\EEA

The equivalent formulation is 
\BEA 
\frac{\p}{\p t}G_{ab}(t,t')=
\sum_c G_{cb}(t,t')\frac{\partial}{\partial m_c}
 \int\d x W(\beta x) {\y_a}p(x|m_1,m_2)
\EEA

The derivation of equal-time responses is a bit tedious.
Let us take $a=b=1$ and change the field from $H$ to $H+\Delta H(k)$
at the time steps $t+k/N$, ($k=1,\cdots, n)$.
It holds that
\BEQ
<M_1(t+\frac{k+1}{N})>-<M_1(t+\frac{k}{N})>=
\int \d y_1\d y_2\d M_1\d M_2\, W(\beta \delta E_k)\,y_1
p(y_1,y_2|M_1,M_2)\,p(M_1,M_2,t+\frac{k}{N})
\EEQ
where
\BEA \label{dE=}
     \delta E_k&=&\frac{K}{2}y_2-(H+\Delta H(k))(M_1+y_1)
+(H+\Delta H(k-1))M_1
\qquad (k=1,\cdots n)
\EEA
Generalizing to all four case we have a response
\BEA\label{Gab(t+t)=}
G_{ab}(t^+,t)&\equiv&\lim_{N\to\infty}  \frac{1}{\Delta t}\,\,
\frac{\partial <m_a(t+\Delta t)>}{\partial \Delta H_b}\nn
&\equiv&
\lim_{N\to\infty}\frac{<m_a(t+\frac{n+1}{N})>-{\rm idem}(\Delta H(k)=0)}
{\d t\,\,\sum_k\Delta H(k)}\nn
&=&G_{ab}^{\rm main}(t^+,t)+G_{ab}^{\rm switch}(t^+,t)\EEA
where we used that $\d t=1/N$ and $\Delta t=n/N$. The main term 
arises from the $y_a$ terms in  (\ref{dE=}),
\BEA G_{ab}^{\rm main}(t^+,t)&=&
-\beta \int \d y_1\d y_2 W'(\beta x)\,y_ay_b p(y_1,y_2|m_1,m_2)\nn
&=&-\beta \int \d x W'(\beta x)(\,{\overline y}_a{\overline y}_b
+\left(-\frac{H_1}{H_2}\right)^{a+b-2}\Delta_y )p(x|m_1,m_2)
\qquad (a,b=1,2)\EEA
where $W'(\beta x)=-\exp(-\beta x)$ for $x>0$ and zero for $x<0$.
The contribution for switching on and off comes from the
$M_a(\Delta H_a(k)-\Delta H_a(k-1))$ terms, 
\BEQ G_{ab}^{\rm switch}(t^+,t)
=\beta \frac{\d}{\d t}\left\{m_b(t)\int \d y_1\d y_2 W'(\beta x)\,y_a
p(y_1,y_2|m_1(t),m_2(t))\right\}
\EEQ
does not depend on the precise switching procedure, but neither
 does it not vanish for adiabatic procedures.
 In our models this term will have contributions proportional to
$\dot m_{1,2}\sim 1/t$, so in the large-$t$ domain it is
much smaller than the terms of our interest.
We shall neglected it from now on. We should point, however,
 out that the responses $G_{m\e}$ and $G_{\e\e}$ do not involve 
such switching terms. This is related to the nature of the 
Monte-Carlo dynamics.

For the responses with respect to an instantanous
 temperature pulse one has 
\BEA G_{m\e }(t^+,t)=
-\beta \int \dx W'(\beta x)xy_0 p(x|\e)\EEA
\BEA \label{Geet+t=}
G_{\e \e}(t^+,t)=
-\beta \int \dx W'(\beta x)x^2 p(x|\e)\EEA
The relation with $G_{ab}$ is exactly as for the $C$'s, see below
(\ref{mambdef}), as these two quantities donot suffer from switching effects.

\section{Glassy dynamics of the oscillator model}
\label	{OscSolution} 
\setcounter{equation}{0}\setcounter{figure}{0} 
\renewcommand{\thesection}{\arabic{section}.}
We are considering a system with one mode.
In view of the equilibrium relation $\e_{eq}=T/2$ we may
introduce the effective temperature $T_e$ by
\BEQ T_e(t)\equiv 2\e(t) \EEQ

The dynamics of the oscillator model
simplifies in the region $T\ll A=K\Delta^2/8$.
Technically this occurs since for $\e\ll A$ or $T_e\ll A$
we can approximate in the expression
\BEQ \label{pxe=}
p(x|\e)=\frac{1}{4\sqrt{\pi AT_e}}
\exp(-\beta_eA+\frac{\beta_ex}{2})
\exp(-\frac{(\beta_ex)^2}{16\beta_eA}) \EEQ 
the last, Gaussian factor 
by $1-x^2/16AT_e$, leaving only exponential integrals.
We shall investigate the dynamics in this region, and look for 
relations satisfied by observables. 

\subsection{Equilibrium regime}
In equilibrium one has $\e=T/2$, $m=H/K$. Then there holds detailed balance
\BEQ W(\beta x)p(x|\e)=W(-\beta x)p(-x|\e)\EEQ
assuring that $\dot\e=0$ in eq. (\ref{dedt=}). 
For eq. (\ref{dDE2dt=}) this implies
\BEQ { \langle\delta E^2\rangle}=\half N T^2 \EEQ
in accordance with the equilibrium  relation 
$\d U/\d T=\beta^2\langle\delta \H^2\rangle$. The relation
(\ref{MMeq}) amounts to
\BEQ { \langle\delta M^2\rangle}=\frac{N T}{K} \EEQ
This is also expected, since  only the diagonal $i=j$ terms in 
$\langle \delta M^2\rangle =\sum_{ij} \langle
\delta x_i\delta x_j\rangle $ contribute at equilibrium,
showing that $K\langle\delta M^2\rangle/2$ will indeed reduce to 
$NT/2$, the equilibrium value of $E$.
Finally, eq. (\ref{EMcorr}) tells that the 
cross-correlation $\langle \delta E\delta M\rangle$ vanishes.

The evolution equation (\ref{dedt=}) for the energy can be expressed
in the notation of ref.~\cite{BPR}
\BEQ\label{dedt}
\frac{\d \e(t)}{\d t}=
-(2\e(t)-\frac{T}{2})f(t)+2A\,\,\erfc(\alpha(t))
\EEQ
where $A=K\Delta^2/8$, $\alpha(t)=\sqrt{A/2 \e(t)}$, $f(t)$ was
defined in eq. (\ref{f=}), and
\BEA\label{erfc=}
\erfc(\alpha)&=&\frac{2}{\sqrt{\pi}}\int_\alpha^\infty \dx e^{-x^2}\\
&\approx&
\frac{e^{-\alpha^2}}{\alpha\sqrt{\pi}}(1-\frac{1}{2\alpha^2}
+\frac{3}{4\alpha^4})
\qquad(\alpha\gg 1)\EEA

We can look at relaxation close to equilibrium, where 
$\alpha_{eq}=\sqrt{\beta A}$. We set $\e=T/2+\delta \e$.
Eq (\ref{dedt=}) becomes to linear order 
\BEQ
\frac{\d \delta \e}{\d t}=-\delta\epsilon\left(
8\beta A(2\beta A+1)\erfc(\sqrt{\beta A})
-16\beta A\sqrt{\frac{\beta A}{\pi}}e^{-\beta A}
\right)\equiv
- \frac{\delta\epsilon}{\tau_{eq}^{(\e)}}
\EEQ
When $T\ll A$  the equilibrium timescale becomes,
due to the expansion (\ref{erfc=}) 
\BEQ\label{taueqe}
\tau_{eq}^{(\e)}\approx
 \sqrt{\frac{\pi\beta A}{64}}\,\,e^{\beta A}\EEQ
The latter also follows from (\ref{dedt=})
by performing the $x$-integral, after neglecting 
the $\exp(-x^2/2\Delta_x)$ factor of $p(x|\e)$.


From (\ref{dmdt=}) it is clear that the magnetization relaxes 
to its equilibrium value $m_{eq}=H/K$ at timescale
\BEQ\label{taueqm}
\tau_{eq}^{(m)}
=\frac{1}{f(\infty)}=\frac{2T}{A}\tau_{eq}^{(\e)}
\EEQ
The important fact for us is that both time scales have an Arrhenius
behavior $\sim\exp(A/T)$ at low $T$. This implies that 
the oscillators, subject to parallel Monte Carlo dynamics,
can easily fall out of equilibrium at low  enough $T$,
and thus exhibit interesting glassy behavior.

\subsection{Cooling procedures and the glassy transition}
\label{coolingHO}

Eq. (\ref{dedt=}) simplifies in the regime $T_e\ll A$. Indeed,
as $x^2\sim T_e^2\ll \Delta_x=8AT_e$, we can neglect
the Gaussian factor $\exp(-x^2/2\Delta_x)$ of $p(x|\e)$
in (\ref{pxe=}). We thus obtain
\BEQ \label{Tedot00}
\dot T_e =-\frac{2T_e^2 } {\sqrt{\pi AT_e}}
(1-\frac{T^2}{(2 T_e-T)^2})e^{-\beta_e A}
\EEQ
The condition $T_e>T/2$ is typically satisfied, since $T_e>T$ in
cooling and aging experiments, and $T_e\to T$ in heating the glass.

Using (\ref{taueqe}) we can write this 
for $T_e$ close to $T$ in the universal form
\BEQ\label{Tedot0}
\dot T_e =\frac{T-T_e}{\tau_{eq}(T_e)}
\EEQ
We can now introduce the inverse function $\tau_{eq}^{-1}(t)$;
in our case (\ref{taueqe}) it reads to leading order  
$\tau_{eq}^{-1}(t)=A/\ln t$.
Let us then consider a non-linear cooling process of the form~\cite{Nhammer}
\BEQ\label{coolTQ} 
T(t)=(1-\Q)T_g+\Q\tau_{eq}^{-1}(t)\approx (1-\Q)T_g 
+\Q\frac{ A}{\ln t}\EEQ
It involves two parameters: the glassy transition temperature $T_g$ 
and the dimensionless cooling speed $\Q$.
A non-linear cooling experiment of this form
could be performed in any system with a quickly
diverging equilibrium timescale.

We first show  that a glassy transition occurs around time scale
$t_g=\exp\beta_g A$, where one has $T(t)\approx T_g$.
The timescale during which the system basically remains 
at temperature $T$ is
\BEQ \tau_{cooling}= \frac{T(t)}{|\dot T(t)|}\sim t\EEQ 
Its ratio to the equilibrium timescale is
\BEQ
\frac{\tau_{cooling}(T(t))}{\tau_{eq}(T(t))}
\sim 
\left(\frac{t_g}{t}\right)^{\Q-1}
 \EEQ
We can discriminate three cases:
a) When $\Q>1$, then for $t\ll t_g$ there is equilibrium at the
instantaneous temperature $T(t)$.  For
$t>t_g$ the instantanous equilibration time $\tau_{eq}$
is larger the cooling timescale $\tau_{cooling}$, and the system
becomes glassy. 
b) For $0<\Q<1$ this process describes
cooling in a glassy state so slowly, that equilibrium is reached 
around time $t_g$. c) Finally, for $\Q<0$ it describes
heating in the glassy state, and equilibrium is reached around time $t_g$.
 
Eq. (\ref{coolTQ}) implies that
\BEQ\label{Tdot}
 \dot T=\frac{\Q}{\tau'_{eq}(T_g-(T_g-T)/\Q)}
\EEQ
This allows us to combine (\ref{Tedot0}) and  (\ref{Tdot}) into the
time-independent form
\BEQ \label{dTedT=}
\frac{\p T_e}{\p T}\fix_H
=\frac{T-T_e}{\Q}\,\frac{\tau'_{eq}(T_g-(T_g-T)/\Q)}
{\tau_{eq}(T_e)}
\EEQ
For $T-T_g\gg T_g^2/A$ one has the equilibrium value $T_e=T$, with
exponentially small corrections. Well below the glass transition
$T-T_g\ll- T_g^2/A$ one has $T_e=T_g-(T_g-T)/\Q$. Due to
eq. (\ref{taueqe}) this implies
that $\tau_{eq}(T_e)\sim \exp\beta_e A\sim t$, so 
$T_e\approx A/\ln t$. As we shall see in next section, this 
is the same behavior as occurs for aging at $T=0$. 
We may conclude that the system then basically has
forgotten its cooling history, and just ages as at 
any low enough temperature. 
Similar behavior was found by Godr\`eche and Luck in the
backgammon model~\cite{GodrecheLuck}.

We may go to dimensionless variables by putting
\BEQ T=T_g+\frac{T_g^2}{A}x; \qquad T_e=T_g+\frac{T_g^2}{A}y\EEQ
and obtain
\BEQ\label{xy-eqn}
 \frac{\d y}{\d x}=\frac{y-x}{\Q}e^{y-x/\Q} \EEQ
This equation is probably universal. Indeed,
it is a small excercise to check that the very same equation 
follows from (\ref{dTedT=}) when $\tau_{eq}$ has a 
Vogel-Tammann-Fulcher-type law 
$\tau_{eq}\sim \exp(A^\gamma(T-T_0)^{-\gamma})$, and a glass transition 
occurs in a narrow range around some $T_g$ with $T_g-T_0\ll A$.

Equation (\ref{xy-eqn})
can be solved analytically for large negative and large positive $x$.
Let us introduce 
\BEQ w=(\frac{1}{\Q}-1)x \EEQ
For large negative $w$ one sets 
\BEQ y=x-\ln z(v);\qquad v=e^w=e^{-(\Q-1)x/\Q} \EEQ
to obtain
\BEQ (\Q-1)v^2z'(v)+\ln z(v)+\Q vz(v)=0\EEQ
By series expansion one finds 
\BEA z&=&1-\Q v+\frac{\Q}{2}(5\Q-2) v^2 
-\frac{\Q}{3}(29\Q^2-27\Q+6)v^3
+\frac{\Q}{24}(1181\Q^3-1812\Q^2+900\Q-144)v^4
\nn
&-&\frac{\Q}{5}(1529\Q^4-3345\Q^3+2690\Q^2-940\Q+120)v^5
+\cdots \EEA
This implies for the specific heat factor
an exponential approach to equilibrium
\BEA \label{CVlarge}
\frac{\p T_e}{\p T}\fix_H &=&\frac{\d y}{\d x}
=-\frac{\ln z(v)}{\Q v z(v)}\\
&=& 1+(\Q-1)\left(-v
+2(2\Q-1)v^2
-\frac{3}{2}(5\Q-2)(3\Q-2)v^3 
+\frac{4}{3}(4\Q-3) (29\Q^2-27\Q+6)v^4\right)+\cdots\nonumber
 \EEA
When $\Q>1$ or $\Q<0$ this applies for large positive $x$.
If $0<\Q<1$ it applies for large negative $x$.
 
For large positive $w$ one sets
\BEQ s=\frac{1}{w}=-\frac{\Q}{(\Q-1)x}\EEQ
and 
\BEQ y=-\frac{1}{(\Q-1)s}+\ln s-\ln u(s) \EEQ
yielding 
\BEQ 
u(s)=1+s\ln s-s\ln u(s)-(\Q-1)su(s)+(\Q-1)s^2 u'(s)
\EEQ
By iteration one finds an expansion in powers of $s$ and $\Lambda=\ln s$
\BEA 
u(s)&=&
1+(\Lambda-\Q+1)s+(2\Q-2-\Lambda)s^2+
(-8\Q+\frac{11}{2}+3\Lambda+\frac{1}{2}\Lambda^2
-2\Lambda\Q+\frac{5}{2}\Q^2)s^3 \nn
&+&(-\frac{57}{2}\Q^2+\frac{16}{3}\Q^3+45\Q+18\Lambda\Q+2\Lambda^2\Q
-5\Lambda\Q^2-\frac{1}{3}\Lambda^3-\frac{7}{2}\Lambda^2-14\Lambda-
\frac{131}{6})s^4
\EEA
This implies for the specific heat factor an algebraic
approach to the value $1/\Q$, with logarithms in the subleading terms
\BEA \label{CVsmall}
\frac{\p T_e}{\p T}\fix_H
&=&\frac{1+s\Lambda-s\ln u(s)}{\Q u(s)}\nn
&=&\frac{1}{\Q} + \frac{\Q - 1}{\Q}\left[\,s + (\Q-\Lambda-2)\,s^{2} 
+(\Q^{2}-2\,\Q\,\Lambda-7\,\Q+\Lambda^{2}+ 7 + 5\,\Lambda)\,s^{3}
 \right.\nn&+&
(2\Q^{3}-35\Q^{2}-6\Q^{2}\Lambda+46\Q\Lambda+92\Q 
+ 6\Lambda ^{2}\Q -52\Lambda-17\Lambda^{2}-61-2\Lambda^{3})\frac{s^{4}}{2}]
+ \cdots 
\EEA
When $\Q>1$ or $\Q<0$ this applies for large negative $x$.
If $0<\Q<1$ it applies for large positive $x$.
It is trivial to see that both (\ref{CVlarge}) and (\ref{CVsmall})
go the correct value $\p T_e/\p T=1$ in the equilibrium limit
$\Q\to 1$. In that limit the system will remain in equilibrium,
because the cooling procedure is very slow.

In figures 4.1-4.4 we present the universal lineshapes for the 
specific heat factor $\d T_e/d T=2\,c$ for several values of $\Q$.  
They exhibit the features known from experiments.

\begin{figure}[htb]
\label{ericR=pos}                  
\epsfxsize=11cm
\centerline{\epsffile{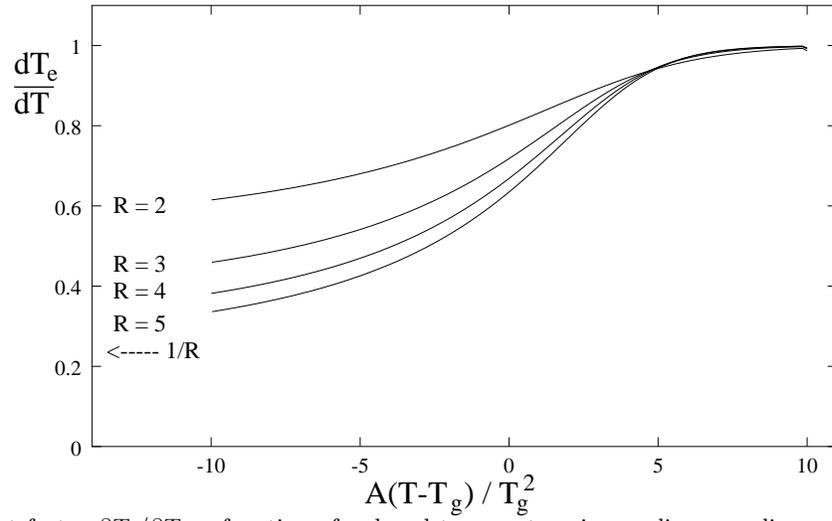 }}
\caption{Specific heat factor $\p T_e/\p T$ as function of reduced
temperature in non-linear cooling experiments with different speed
parameter $\Q$. The asymptotic values are $1$ to the right and $1/\Q$
to the left.}
\end{figure}

\begin{figure}[htb]
\label{ericR=neg}
\epsfxsize=11cm
\centerline{\epsffile{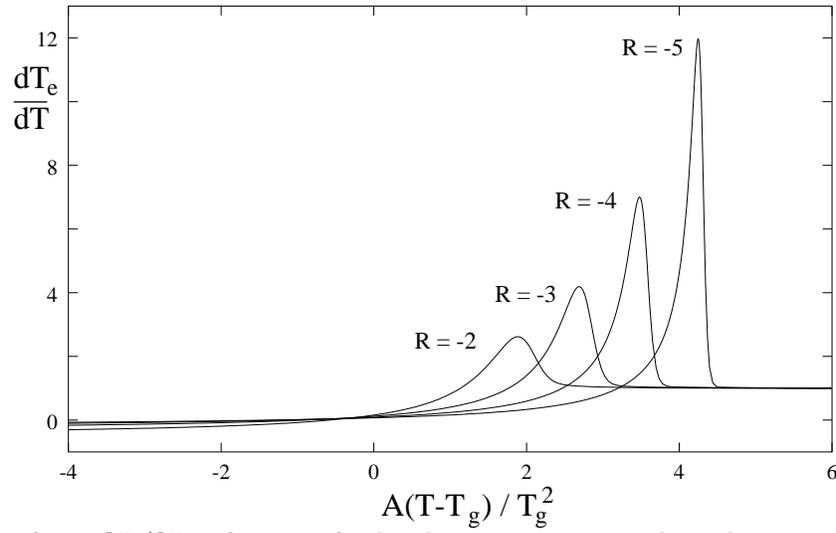}}
\caption{Specific heat factor $\p T_e/\p T$ as function of reduced
temperature in non-linear heating experiments with different speed
parameter $\Q$}
\end{figure}

\begin{figure}[htb]
\label{ericR=pm2}
\epsfxsize=11cm
\centerline{\epsffile{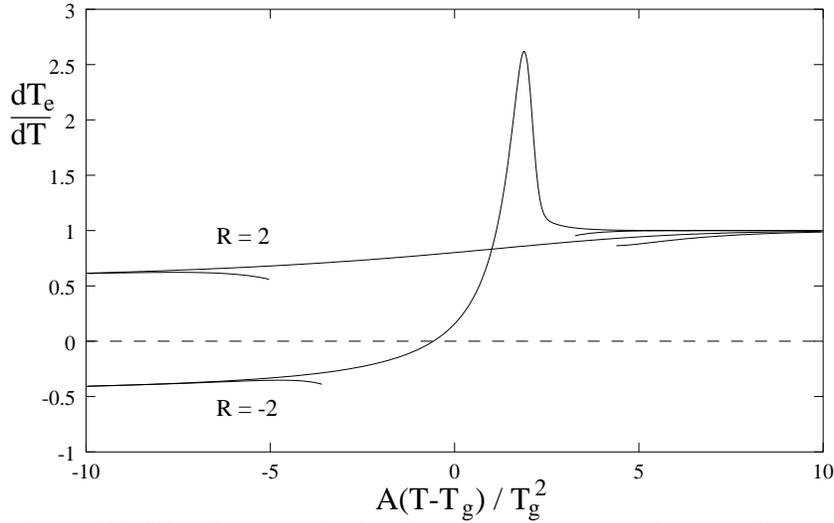}}
\caption{Specific heat factor $\p T_e/\p T$ as function of reduced
temperature in a non-linear cooling experiment with $\Q=2$ and in a
non-linear heating experiment with $\Q=-2$. Dashed lines are
asymptotes from eqs. (\ref{CVlarge}) and (\ref{CVsmall}).}
\end{figure}

\begin{figure}[htb]
\label{ericfig4}
\epsfxsize=11cm
\centerline{\epsffile{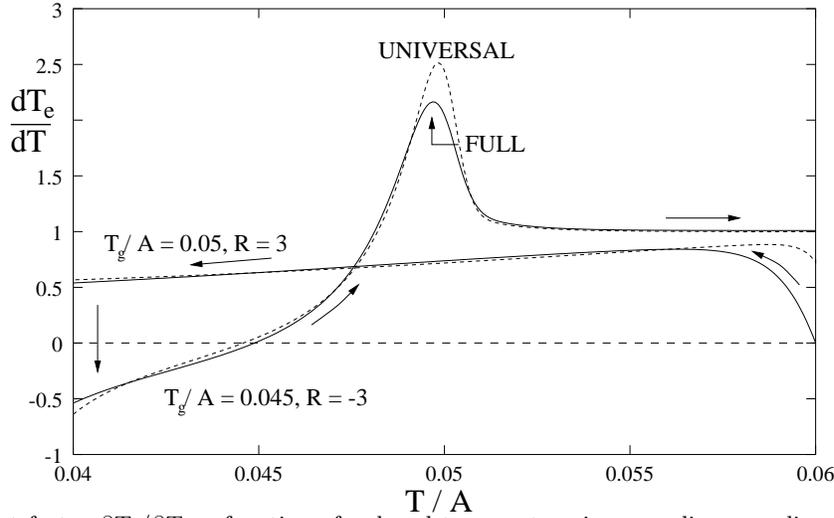}}
\caption{Specific heat factor $\p T_e/\p T$ as function of reduced
temperature in a non-linear cooling experiment with $\Q=3$,
and for a non-linear heating experiment with $\Q=-3$.
The ``universal'' curve is from previous theory. In the ``real model''
the original differentio-integral equations have been used, and
$R$ has changed from $R=3$ to $R=-3$ after reaching $T=0.045A$.
}
\end{figure}

The analysis of this section thus
shows that cooling in systems with an Arrhenius law leads to glassy
behavior quite similar to that expected for realistic glasses.

\subsection{Aging in the glassy regime}

Suppose we quench the system  at time $t=0$ from an equilibrium state at 
temperature $T_{quench}\ll A$ to a lower temperature $T$.
As opposed to previous section, we now assume that after the
quench the system is far from equilibrium 
(viz. $T_{quench}-T\gg T_{quench}^2/A$).
Then eq. (\ref{Tedot0}) may be written as
\BEQ \label{Tedot}
\frac{\d (e^{\beta_eA})}{
(1-r^2)\sqrt{\beta_eA}}=
\frac{2}{\sqrt{\pi}}\d t
\EEQ
where we use the short hand
\BEQ r=\frac{T}{2T_e-T} \EEQ
Its integral is
\BEQ
\frac{e^{\beta_eA}}{(1-r^2)\sqrt{\beta_eA}}
=\frac{2}{\sqrt{\pi}}(t+t_0)
\EEQ
where $t_0$ follows by inserting at $t=0$ the value $T_e=T_{quench}$.
This result may be written as
\BEQ\label{betae=}
\beta_eA -\frac{1}{2}\ln{\beta_eA}
=\ln \frac{t+t_0}{\tau_0};
\qquad \tau_0=\frac{\sqrt{\pi}}{2( 1-r^2)}
\EEQ
For large $t$ one has to leading order $T_e=A/ \ln t$, while 
the initial condition gives
a small correction of order $t_0/t\sim \exp(-\beta_eA)$; it 
may thus be neglected for $t> 10 t_0$. 
This says the initial condition is washed out, 
and is the basis for our interpretation that each decade is
practically  independent of the previous one.
Likewise, the effect of a finite $T$ is very small, and to
leading order one could set $T=0$. This says that in the glassy regime
the energy essentially
evolves as if the system had been quenched to $T=0$.
Only near the return to equilibrium the $T$-dependent factor
in (\ref{betae=}) brings a vanishing argument in the logarithm, from
which non-trivial behavior results, as discussed in section
~\ref{coolingHO}.

To leading order one may invert eq. (\ref {betae=}), to obtain
\BEQ \label{Teaging}
T_e(t)  \approx \frac{A}
{\ln\frac{t}{\tau_0}+\frac{1}{2}\ln \ln\frac{t}{\tau_0}}
\EEQ 
In practice this need not be a good approximation since $1/\ln t$
is usually not very small.
For our purposes (leading order expansion in powers of $T_e$) 
this is equivalent to 
$A/(\ln t+\frac{1}{2}\ln\ln t)$, and actually even to $A/\ln t$. 
It is a simple excercise to check that one has 
\BEQ  \tau_{eq}^{(\e)}(T_e)  \approx
(1-\frac{T^2}{(2T_e-T)^2})\frac{2t}{\sqrt{\pi}} \sim t \EEQ
proving our general assertion that in the aging regime
the effective temperature
also follows by equating the equilibrium time scale to $t$.

In the aging regime eq. (\ref{f=}) becomes 
\BEQ f=2\sqrt{\frac{\beta_eA}{2}}e^{-\beta_eA}(1+r)(1-\frac{T_er^2}{2A})
=\frac{1}{(1-r)t}=\frac{2T_e-T}{2(T_e-T)t}
\label{fres}
\EEQ
Let us define its integral as
\BEQ h(t)={\rm const}\times\exp\int_0^t \d t' f(t') \EEQ
Using $\d \ln h/\d \e=f/\dot\e$ we obtain
\BEQ \label{ht=}
h(t)=
(1-\frac{T\beta_e}{2})
\frac{(\beta_eA)^{3/2} e^{\beta_eA/2}}
{(1-T\beta_e)^{\beta A/2+5/4}}
\EEQ
For $T>0$ this behaves as $\sqrt{t}$, with logarithmic corrections.
At $T=0$ the resulting asymptotic scaling $h(t)\sim t(\ln t)^2$
 differs from the result $h\sim t\sqrt{\ln t}$ reported by ~\cite{BPR}.
However, their figure 2 already show a deviation between the 
data and their asymptotic formula, that becomes increasingly worse
in the asymptotic limit. We were informed 
by E. Hennes that the present expressions (\ref{ht=}), (\ref{betae=})
give for $t'\ge 100$ 
an almost perfect agreement with the numerical solution of the 
integro-differential equations~\cite{EHennes} .

For later use, we mention the results 
\BEQ g=2e^{-\beta_eA}\sqrt{\frac{\beta_eA}{\pi}}(1-r^2+
\frac{T_e}{2A}r^2(1+r)(1+3r))
\EEQ
and 
\BEQ\label{tildeht=} 
\tilde h(t)={\rm const}\times\exp\int_0^t \d t' g(t')
=(\beta_eA)^{3/2}\frac{(1-T\beta_e/2)^2}{1-T\beta_e}\,e^{\beta_eA}
 \EEQ

Due to eq. (\ref{dmdt=}), the magnetization relaxes as
\BEQ m(t)=\frac{H}{K}+(m(t_0)-\frac{H}{K})\frac{h(t_0)}{h(t)}
\EEQ
In the regime of large times and small $T_e\sim 1/\ln t$, the deviation 
$m(t)$ from $H/K$ is exponentially small in $T_e$. As compared to
the powerlaw that occurs in the energy (recall that $\e=T_e/2$), 
this can be neglected.
This says that the magnetization quickly goes to its quasi-stationary
value.

\subsection{Correlations}

Bonilla et al.  considered the on-site correlation function
$\langle x_i(t)x_i(t')\rangle$~\cite{BPR}.  At non-zero field it would
become $\langle \delta x_i(t)\delta x_i(t')\rangle$, 
with $\delta x_i=x_i-<\!x_i\!>(t)$.
However, the correlation function related to thermodynamics 
is the global correlator
$C_{mm}(t,t')=\sum_{i,j}\langle \delta x_i(t)\delta x_j(t')\rangle$, 
defined in (\ref{mmdef}). Its equal time value is found from 
eq. (\ref{MMeq}). We shall now study its two-time structure.

 Let us introduce
\BEQ
\xav\,=\frac{\int \d x W(\beta x) x p(x|\e)}{\int \d x W(\beta x) p(x|\e)}
\approx -4T_e\frac{T_e-T}{2T_e-T}=-2T_e(1-r)
\EEQ where the explicit result holds for $T_e$ and $T$ much smaller
than $A$.
After dividing by (\ref{dedt=}) we may write (\ref{MMeq}) as
\BEQ\label{dM^2osc}
 2\xav\frac{\dot C_{mm}(t,t)}{\dot T_e(t)}= 
\frac{C_{mm}(t,t)}{T_e(t)}(2\xav-K\Delta^2)+\Delta^2
\EEQ 
where we used that $<\!\delta E\delta M\!>$ is exponentially small in $T_e$.
The dominant behavior follows by neglecting the $\xav\sim T_e$ terms. 
However, they can be fully taken  into account, 
as the solution to this equation reads
\BEQ \label{Cmmtt=}
C_{mm}(t,t)=\frac{T_e(t)}{K}\EEQ
Corrections are exponentially small in $T_e$.
This result even holds when $T$, occurring in $\xav$, depends on time.
It also allows a simple check: The result 
$K\langle\delta M^2\rangle/2$$\approx$$E=NT_e/2$ is
in accordance with the expectation of Bonilla et al ~\cite{BPR}
that off-diagonal terms $\langle \delta x_i \delta x_j\rangle$ with 
$i\neq j$ are subleading. 

Solving eq. (\ref{dCmmdt=}) 
for $t'\neq t$ we can neglect the $C_{\e m}$ term, since 
it is exponentially small in $T_e$. This yields
\BEQ \label{Cmmttp=}
C_{mm}(t,t')= \frac{T_e(t')}{K}\,\frac{h(t')}{h(t)}\EEQ
where $h(t)$ was defined in eq. (\ref{ht=}).

\subsubsection{Stretched exponential fitting 
and arguments against doing that}

In the study of glasses, where often at best two orders of magnitude
of $C$ can be determined,  it is commonly assumed that there occurs a
stretched exponential  decay,
\BEQ C_{mm}(t,t')=a(t')\exp(-(t/\tau)^\gamma) \EEQ
It is often stated that such stretched exponential decay is one of the
basic properties of the glassy state.

In our case we would need that 
\BEQ \label{strexp}
h(t)=\exp(-d+(t/\tau)^\gamma)\EEQ
for some set of parameters $d$, $\tau$, $\gamma$, or, equivalently, that
$\ln(d+\ln h(t))$ is linear in $\ln t$ with slope $\gamma$.
In view of the exact expression, this is certainly not 
an exact description. Let us, however, look at the plot for $d=0$ in figure
4.5. 

\begin{figure}[htb]
\label{stretchedexpfig}
\epsfxsize=9cm
\centerline{\epsffile{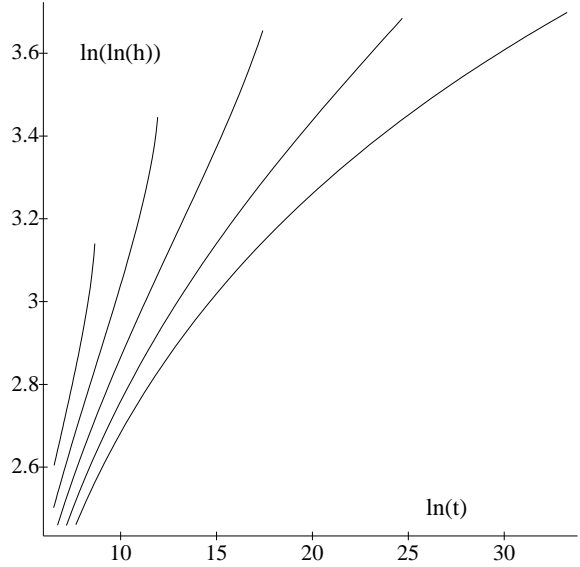}}
\caption{The decay of the correlation function can be described
by a  stretched exponential in a time-window 
where the data for $\ln\ln h$ are linear in
$\ln t$. The stretching exponent then equals the slope in this figure,
and will depend on $T$ and on the chosen time-window.
From left to right: $T/A=0.1$, $0.075$, $0.05$, $0.025$, $0$.
The bending of in the lines on the left (having a relatively large $T$)
 indicates that equilibrium is approached at the considered timescale.}
\end{figure}

It is seen that $\ln\ln h(t)=\gamma \ln t -\gamma \ln \tau$
can be a reasonable approximation in a not-too-wide large-time window 
$t_{min}< t < t_{max}$. In agreement with usual findings, 
the effective exponent $\gamma$ will decrease with $T$, and be bounded by the 
finite $T=0$ value. Notice, however, that it will also depend on the
time window where the fit is made.

In the stretched exponential fitting procedure there is one more adjustable
parameter, namely the overall prefactor $\exp(-d)$. In figure 4.6 
we take $T=0.0025\,A$ and give plots
of $\ln(d+\ln h)$ versus $\ln t$ for various $d$. 
In intervals where this curve is 
flat, $h$ is well described by a stretched exponential (\ref{strexp}).
\begin{figure}[htb]
\label{lhdfitplot}
\epsfxsize=11cm
\centerline{\epsffile{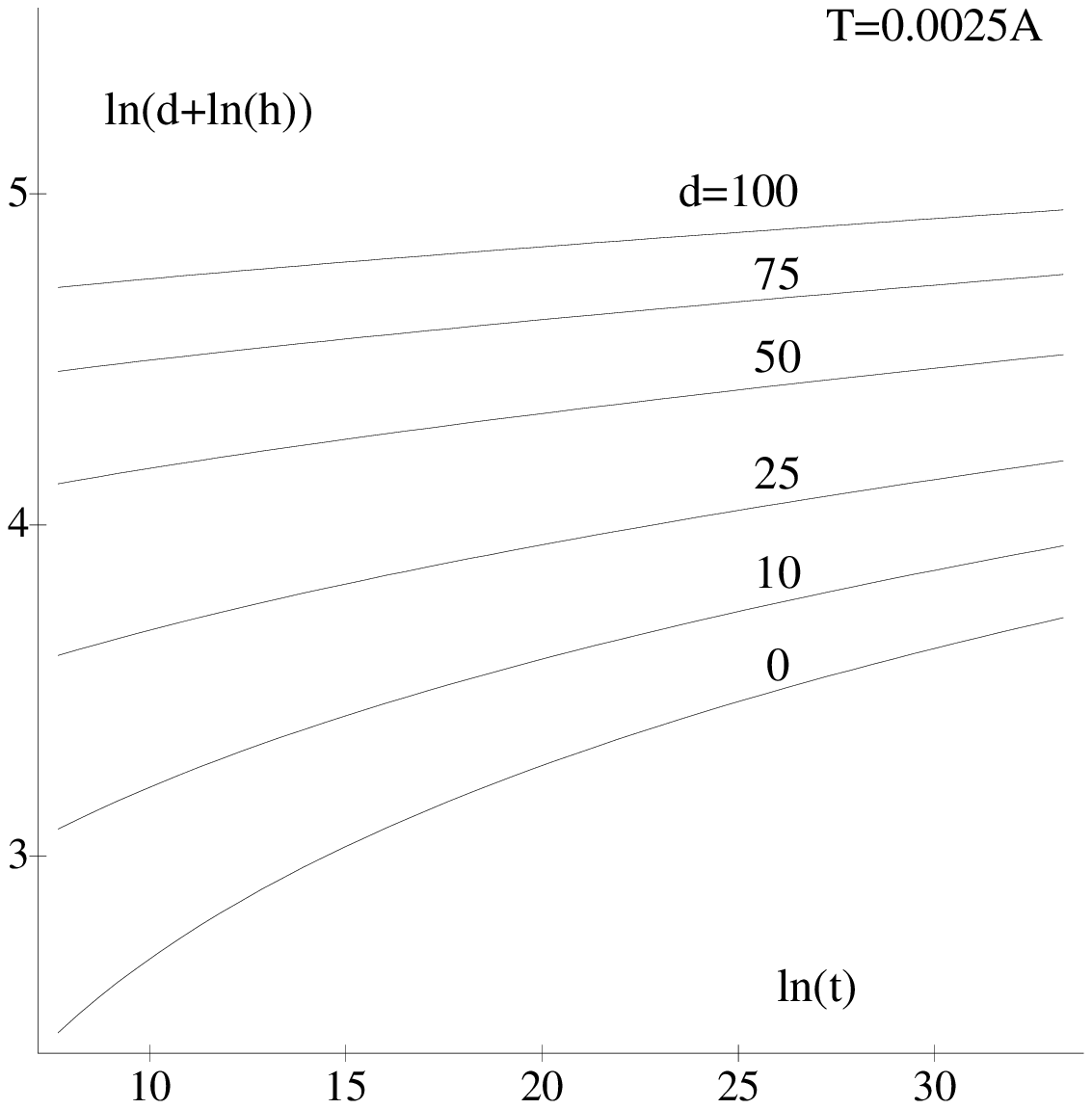}}
\caption{Log-log plots of the function $d+\ln h(t)$  at $T=0.0025\,A$ 
for various $d$. $h$ is well described by a stretched exponential 
$h(t)\sim\exp(-d+(t/\tau)^\gamma)$ in an interval 
where one of the plotted lines is straight. 
Then the slope yields $\gamma$ and the offset $-\gamma\ln\tau$. }
\end{figure}
This information can be used to obtain $C(t,t')\sim h(t')/h(t)$.
To exagerate what happens, we take a very simple linear fit
to the data of figure 4.6: we consider the interval
$10\le \ln t\le 20$ and make, for a given value of $d$, 
a linear interpolation through the data
points at $\ln t=10$ and $\ln t=20$. Taking $t'=\exp(10)$
we plot in figure 4.7 the fits to
$\log (h(t')/h(t))$ for the cases $d=10$, $d=100$, and compare
with the exact result $h(t')/h(t)$ from eq. (\ref{ht=}). 
By our construction, the results agree at
$\ln t=10$ and $20$. It is seen that in all cases the fits are
reasonable in regard of the scale presented in the figure, and
that increasing  $d$ improves the overall fit. 

\begin{figure}[htb]
\label{hthtpfeps}
\epsfxsize=11cm
\centerline{\epsffile{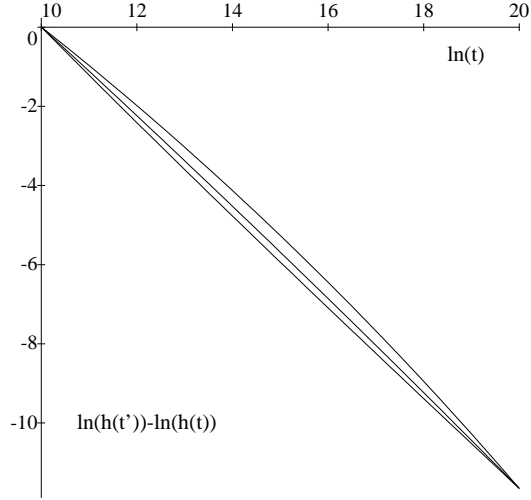}}
\caption{Log-log plot of stretched exponential fits to the
ratio $h(t')/h(t)$  at fixed $t'=\exp(10)$ and $T=0.0025\,A$.
Upper line: by taking a linear interpolation between the points 
$\ln t=10$ and $\ln t=20$ for the case $d=10$ of previous figure.
Middle line: the same for $d=100$.
Lower line: exact result.}
\end{figure}

The free fitting parameter $d$ is not present in reality.
It occurs if one overlooks that $h(t)/h(t')$ should be fitted
as function of {\it two} parameters, namely both $t$ and $t'$.
Indeed, at one given $t'$ one is  free to choose $d$; however,
emposing the asymmetry $t\to t'$ would bring $d\to (t'/\tau)^\gamma$.
This reduced freedom would 
decrease the overall accuracy of the fit. In practice this
is typically not done, partly because of lack of data curves 
$C(t,t')$ at enough $t'$ values. Nevertheless, whenever fitting 
of stretched exponentials is attempted, we stress to make a 
two-parameter fit of $C(t,t')$.

In our opinion the present fitting procedures merely say that stretched
exponential fits can too often be made, without yielding too much insight.
The standard statement that stretched exponential decay is one of the
basic properties of the glassy state should, in our view, 
be taken with a big grain of salt. What really happens is a slow decay,
of which too few orders of magnitude are known to draw firm
conclusions on their analytical form. It seems needless to say that
in experiments the tails of the correlations have large relative
errors, which, in our opinion, make the problem quite
insensitive to stretched exponential or many other fitting procedures.

We feel that the situation even becomes worse if such fitting is
applied for showing the presence of  critical behavior near a 
supposed critical temperature in the glass, as is sometimes done.

\subsection{Fluctuation-dissipation relations}

We now consider aging dynamics at fixed $T\ll A$.
Neglecting all $y_0$ contributions, we find from (\ref{Gab(t+t)=})
the equal time value $G_{mm}(t,t)=f(t)/K$. This result is exact
for $H=0$~\cite{BPR}. Its two-time form 
follows from (\ref{Gmmttp=})
\BEQ G_{mm}(t,t')= \frac{f(t')}{K}\frac{h(t')}{h(t)}\EEQ
We can now consider the fluctuation-dissipation relation.
We define the effective temperature $\tilde T_e$ by
\BEQ \frac{\partial_{t'}C_{mm}(t,t')}{G_{mm}(t,t')}
=\tilde T_e(t')\EEQ
By direct evaluation of the left hand side,
we find from (\ref{Cmmttp=}), using (\ref{Tedot00}) and
(\ref{fres}),  
\BEQ\label{tildeTe=}
\tilde T_e(t)=T_e(t)+\frac{\dot T_e(t)}{f(t)}
=T_e-\frac{T_e^2}{A}\,\frac{2(T_e-T)}{2T_e-T}+\O(\frac{T_e^3}{A^2})
\EEQ
Due to eq. (\ref{dCmmdt=}) this agrees with the general result
(\ref{tildeTegen}), which assures consistency with
single-time expressions.
As the second term is smaller by a factor $T_e/A$ with respect to $T_e$,
we thus see that to leading order the same effective temperature 
occurs as in the  energy and the time-scale. (The same happens
when the local correlator is considered ~\cite{BPR}).
Notice, however, that the leading correction is non-universal, 
as it depends on the model parameter $A$.
When equilibrium is approached,  $T_e\to T$, and  the
correct limit $\tilde T_e=T$ is reproduced.

Let us now look at the energy fluctuations. To leading order we 
may neglect $\d\langle \delta E^2\rangle/\d t$ in eq. (\ref{dDE2dt=}).
This brings 
\BEQ \label{explEfluct}
\frac{1}{N}\,<\! \delta E^2\!>=C_{\e\e}(t,t)
=\frac{-T_e^2<\! x^2\!>}
{4A\xav}\approx \frac{T_e^3(t)}{A}\,
\frac{1+r^3}{1-r^2}\EEQ
This result shows various points. First, even at $r=T=0$ it is one
order of magnitude smaller than what one would
anticipate from the equilibrium expression $C_{\e\e}=T^2/2$.
Only in the glass transition region $T_e-T\sim (1-r)T\sim T^2/A$ 
the equilibrium scaling $C_{\e\e}\sim T^2$ is recovered.
 We conclude that there is a complicated, non-universal
$T$-dependence in the whole aging regime $T<T_e$. 
The possibility of a model-independent generalization outside
equilibrium of the relation 
$\d U/\d T=\beta^2\langle\delta E^2\rangle$ will be discussed in 
the Discussion.

For different times we find from (\ref{dCeedt=}), (\ref{tildeht=})
\BEQ
C_{\e\e}(t,t') =\frac{T_e^3(t')}{A}\,
\frac{1+r^3(t')}{1-r^2(t')}\,\frac{\tilde h(t')}{\tilde h(t)}
\EEQ
Likewise, the Greens function follows from (\ref{Geet+t=}) 
and (\ref{dGeedt=}) as 
\BEQ G_{\e\e}(t,t')=\frac{\tilde h(t')}{\tilde h(t)}
\left(\frac{2T_e^2r^2(1+r)e^{-\beta_e A}}{\sqrt{\pi A
T_e}}\right)(t')
\EEQ
To leading order this yields the fluctuation-dissipation relation
\BEQ\label{Teeefdef}
 \frac{\partial_{t'}C_{\e\e}(t,t')}{G_{\e\e}(t,t')}
=T_e^{(\e\e)}(t')\EEQ
where the $t$-dependence again has dropped out. The quantity 
\BEQ \label{Teeef=}
T_e^{(\e\e)}(t)
=\frac{T_e(t)(1+r^3(t))}{r^2(t)(1+r(t))}
=\frac{T_e(4T_e^2-6T_eT+3T^2)}{T^2}
\EEQ
is also an effective
temperature that also has the correct limit when $T_e\to T$. 
Unlike $T_e$ itself, it has no obvious model-independent 
interpretation. We feel that this is due the fact 
that it relates to subleading quantities.

For the  fluctuations (\ref{Cmmtt=}) in $M$ the 
non-universal terms are exponentially small in $T_e$.
This shows that for the fluctuations in $M=M_1$ a quasi-universal
behavior takes place. As $C_{\e m}$ is negligible, the 
$M_1M_2$ cross-fluctuations are also simple,
\BEQ C_{12}(t,t')=\frac{2HT_e(t')h(t')}{K^2h(t)}\EEQ
plus exponentially small corrections in $T_e$, or powerlaw in $1/t$.
The $M_2$ correlations and responses have the form
\BEQ
C_{{22}}(t,t')=\frac{4H^2}{K^2}C_{11}(t,t')+
\frac{4}{K^2}C_{\e\e}(t,t'); \qquad
G_{22}(t,t')=\frac{4H^2}{K^2}G_{11}(t,t')+\frac{4}{K^2}G_{\e\e}(t,t')
\EEQ
In both expressions the first term is two orders of magnitude
in $T_e$ larger. On top of that,
the second term decays faster ($\sim t'/t$ versus $\sqrt{t'/t}$) 
whenever $T$ is non-zero. 
It thus holds that in all four cases ~\cite{Nhammer}
\BEQ \frac{\partial_{t'}C_{ab}(t,t')}{G_{ab}(t,t')} 
=\tilde T_e(t')\qquad (a,b=1,2)\EEQ
This simple result suggests that in general the fields could also stand for 
a chemical potential, a pressure, or a quenched randomly directed 
forcing strength.

\subsection{Non-equilibrium thermodynamics}

We now wish to view previous results in the thermodynamic framework
of Section II.

Since there is only one type of processes, that are 
by definition the slow ones,
the entropy of equilibrium processes $S_{\rm ep}$ vanishes. 
For such cases the configurational entropy can be derived simply.
It is defined by the degeneracy of states with energy $U$, and given 
by the micro-canonical partition sum
\BEQ e^\I=\int Dx \,\delta(\H(x)-U)=\int_{-i\infty}^{i\infty}
\frac{ \d\tilde\beta}{2\pi i} 
\, e^{\tilde\beta U}\int Dx\,
e^{-\tilde\beta \H(x)}
\EEQ
where $Dx=\Pi_i\d x_i$ is the integration measure.
By the saddle point method one obtains
\BEQ \I=\max_{\tilde T}\,\,\tilde \beta(U-F_{eq}(\tilde T))
=S_{eq}(T_e)\EEQ 
where we used that for $U=U_{eq}(T_e)$ the minimum is assumed
at $\tilde T=T_e$. This result holds generally 
in simple systems with only one timescale that diverges
near $T=0$. Here we have
\BEQ \I=S_{eq}(T_e)= \frac{N}{2}(\log\frac{ T_e}{K}+1)\EEQ
Since $\d U=N\d T_e/2-(NH/K)\d H$, 
it is now clear that the formulation (\ref{dU=})  of the first law 
is satisfied in
the present non-equilibrium state. As $\S=0$, the free energy reads
\BEQ F=U-T_e \I \EEQ
and it satisfies the relations (\ref{dF=}).

As $m_1=H/K$ is temperature-independent, the modified 
Maxwell relation reduces to the standard one: in eq. (\ref{modMaxH}) 
the terms proportional to $T$ vanish, and the other terms
follow already from (\ref{dU=}) with $\S=0$. Neither is it interesting to
investigate the first Ehrenfest relation (\ref{Ehren1H}):
It holds trivially, as 
one has $\alpha=0$, $\chi=1/K$, implying $\Delta\alpha=\Delta\chi=0$. 
Notice, however, the present results already require that
second Ehrenfest relation (\ref{Ehren2pure})
is  modified outside equilibrium~\cite{NEhren}. 
Indeed, from section \ref{coolingHO} we have
$\Delta C=N(\Q-1)/(2\Q)\neq 0$, while $\Delta \alpha=0$.
Equation (\ref{modEhren2H}) is nevertheless satisfied,
 since $\p T_e/\p T=1/\Q$ and $\d \I/\d T=\p \I/\p T=N/(2T_g)$

The fluctuation formula (\ref{chifluct}) are also satisfied.
To show this explicitly, let us take $a=b=1$. Since there are no fast
processes, the first term vanishes. 
The same holds for  the third term, since $m=H/K$ 
leads to $\p m/\p T_e=0$. Due to eq. (\ref{Cmmtt=})
the second term equals $(NT_e/K)/(NT_e)=1/K$,
which is the desired result. We can also check it by 
integrating the up the instantaneous field pulses,
as was done more generally 
in the argument starting with eq. (\ref{Mabhelp14}). 
The same conclusions hold for the other three cases. 
We have already mentioned that the fluctuation-dissipation
relation (\ref{FDR=}) is satisfied with $\tilde T_e$
 given in eq. (\ref{tildeTe=}), and that the 
apparent specific heat $C_H=N(\p T_e/\p T)/2$ has no simple
connection with the energy fluctuations (\ref{explEfluct}).

In all situations considered we have seen that $T_e\sim A/\ln t$ 
is to leading
order in agreement with the timescale relation $\tau_{eq}(T_e)\sim t$.
We have also seen that correlation function has the scaling $h(t')/h(t)$,
with, at finite $T$,  $h(t)\sim \sqrt{t}$ times a function of $\ln t$.  
In the $T\to 0$ limit they become so strong that they replace 
the $h(t)\sim \sqrt{t}$ scaling by 
$h=\, t\, \times{\rm function}(\ln t)$

In conclusion, the proposed picture applies to the harmonic
oscillator model, be it that a few aspects are quite trivial.
In next section we shall consider
a model of spherical spins, which has a  richer behavior 
when changing field, and in regard to the Ehrenfest relations.

\section{Monte Carlo dynamics of free spherical spins in a random field}
\label{SpherSpins}
\setcounter{equation}{0}\setcounter{figure}{0} 
\renewcommand{\thesection}{\arabic{section}.}

Previous model had the drawback that the effect of a field was rather
trivial. It was therefore of no great interest
to check the first Ehrenfest relation: it is satisfied in a trivial way,
having $\Delta\alpha=\Delta\chi=0$.

We have therefore considered a closely related model, 
containing free spherical spins
in the presence of a random external field, which does not share
these drawbacks~\cite{Nhammer}. Also in this very 
simple model Monte Carlo dynamics can be solved exactly and 
leads to glassy behavior. In fact, the dynamics just maps to
leading order onto the one of the oscillator model of previous section.

 The Hamiltonian contains two parts, 
a ``self-interaction'' term involving
fields $\Gamma_i$, and a coupling to an external field $H$
\BEQ \H=-\sum_{i=1}^N \Gamma_i S_i-H\sum_{i=1}^N S_i\EEQ
The model is solvable for any set of quenched random fields 
$\Gamma_i$ that have average zero and variance $\Gamma^2$. 
To simplify the discussion, we make the additional 
(but technically unnneeded) assumption that
$\Gamma_i=\pm \Gamma$, implying that at each spin position
there is a quenched random unit vector $\Gamma_i/\Gamma$, along 
which the spins wish to point for large pinning field $\Gamma$.
This limitation allows the exact gauge transformation
$S_i\to \Gamma_i S_i/\Gamma$, which interchanges the role of $H$ 
and $\Gamma$. Without the additional assumption this interchange
would also exist; this is due to the spherical nature of the spins.

In terms of the ``staggered''
magnetization $M_s\equiv (1/\Gamma)\sum_i \Gamma_i S_i$
one simply has $\H=-\Gamma M_s-HM$.
When defining $H_1=H$, $H_2=\Gamma$, $M_1=M$, $M_2=M_s$,
we may also write this as 
\BEQ \label{H12=}
\H=-\sum_{c=1}^2 H_cM_c \EEQ

The spins are spherical, which means that they can 
take all real values compatible with
\BEQ\label{spherconstr}
\sum_iS_i^2=N
\EEQ


In equilibrium the system has  a free energy 
\BEQ 
\frac{F_{eq}}{N}=\frac{T}{2}\log\beta\mu-\frac{K^2}{2\mu}-\frac{\mu}{2}
\EEQ
where
\BEQ \label{K=}
K=\sqrt{\Gamma^2+H^2}\EEQ
and with chemical potential $\mu$ fixed by optimization, implying
\BEQ \mu=\sqrt{K^2+\frac{1}{4}T^2}+\half T\EEQ
This yields for the internal energy, the magnetization, and for the entropy
\BEQ 
\dt{U_{eq}} 
=-\frac{K^2}{\mu}=-\sqrt{K^2+\frac{1}{4}T^2}+\half T
\approx -K+\half T-\frac{T^2}{8K}
\EEQ
\BEQ
\dt{M_{eq}} 
=\frac{H}{\mu}\approx \frac{H}{K}-\frac{HT}{2K^2}+\frac{HT^2}{8K^3}
\EEQ
\BEQ \label{SeqS=}
\dt S_{eq}=\half\ln\frac{T}{\mu}+\half \approx \half\ln\frac{T}{K}+\half 
\EEQ
The approximations hold for low $T$.

\subsection{Monte Carlo dynamics}

As for the oscillators, one makes parallel
Monte Carlo moves $S_i\to S_i'=S_i+r_i/\sqrt{N}$,
with the $r_i$ independently drawn from a Gaussian with
average zero and variance $\Delta^2$. 
Next  one makes a global rescaling of the 
length of the spins, to reinforce the spherical constraint.
This leads to the final update per time step
\BEQ 
S_i'=S_i+\frac{r_i}{\sqrt{N}}-S_i\sum_j(\frac{r_jS_j}{N\sqrt{N}}
+\frac{r_j^2}{2N^2})+\cdots
\EEQ
This conserves the constraint (\ref{spherconstr}). 
It implies for the change in the energy and in the
total magnetization $M=\sum_iS_i$
\BEQ
\H'-\H=\sum_i\{-\frac{(\Gamma_i+H)r_i}{\sqrt{N}}
-\H\frac{r_iS_i}{N\sqrt{N}}
-\H\frac{r_i^2}{2N^2}\};\qquad
M'- M=\sum_i\{ \frac{r_i}{\sqrt{N}}-M\frac{r_iS_i}{N\sqrt{N}}-
M\frac{r_i^2}{2N^2}\}
\EEQ
Introducing the new variables
\BEQ \label{e=mu1=}
\e=K+\frac{\H}{N}; \qquad m=\frac{M}{N};\qquad
\mu_1=m_{eq}(\e)-m=-m+\frac{H}{K}-\frac{H}{K^2}\e,
\EEQ
which are small near equilibrium,
this leads to moves $\H'=\H+x$, $M'=M+y$ with
Gaussian transition probabilities of the type 
(\ref{Pxyem=}), (\ref{PxPy=}), having parameters
\BEA
&x_0=\half\Delta^2(K-\e);\qquad 
&\Delta_x=\Delta^2\e(2K-\e) \EEA \BEQ
y_0=-\frac{H}{K^2}x+\mu_1\frac{\Delta^2K^2-2Kx+2x\e}{2\e(2K-\e)}
;\qquad
\Delta_y=\Delta^2\large(\frac{\Gamma^2}{K^2}
-\frac{K^2 \mu_1^2}{\e(2K-\e)}\large)
\EEQ
where
\BEQ K=\sqrt{\Gamma^2+H^2}\EEQ
In particular at small $\e$ the model of spherical spins in a 
random external field leads to a
problem very similar to that of the uncoupled, identical
oscillators in a steady field. The previous general formulae
for variances, correlation- and response functions remain valid here.

\subsection{Glassy transition and the Ehrenfest relations}
\label{coolingS}
The evolution for $\e$ again satisfies eq. (\ref{dedt=}), with 
the new expressions for $x_0$ and $\Delta_x$.
We can again introduce $T_e$ by equating $U=-K+\e=U_{eq}(T_e)$, so that
\BEQ \label{Tedef}
T_e=\frac{(2K-\e)\e}{K-\e}\approx 2\e(1+\frac{\e}{2K});\qquad
\e=K+\frac{T_e}{2}-\sqrt{K^2+\frac{T_e^2}{4}}\approx
\frac{T_e}{2}-\frac{T_e^2}{8K}
\EEQ
We also define 
\BEQ A=\frac{\Delta^2K}{8} \qquad B=\frac{\Delta^2}{8}(K-\e)\EEQ
 $p(x|\e)$ takes the form (\ref{pxe=}) with $A\to B$.
For $T_e\ll B$ one can again approximate  it
by an exponential. This yields the equilibrium timescale
\BEQ
\tau_{eq}^{(\e)}=\frac{\sqrt{K^2+T^2/4}-T/2}{\sqrt{K^2+T^2/4}}
\, \sqrt{\frac{\pi \beta B}{64}}e^{\beta B}
\EEQ
For a non-linear cooling process of the form (\ref{coolTQ})
the results of section \ref{coolingHO} apply immediately.
We consider cooling sequences with $\Q>1$, where the system goes
from a paramagnet to a glassy state in a region around
some $T_g\ll B$.
Below the glassy transition one has an apparent specific heat
\BEQ
C=C_2 \frac{\p T_e}{\p T}\fix_H 
\EEQ
This is of the general form (\ref{CpTool}), with background $C_1=0$,
since there are no fast processes in the present model.
It holds that 
\BEQ
c_2=\frac{C_2}{N}=\frac{1}{2}-\frac{T_e}{4\sqrt{K^2+T_e^2/4}}\approx 
\frac{1}{2}-\frac{T_e}{4K} \EEQ
The same universal lineshapes of the oscillator
 model thus occur here, with non-trivial prefactor $c_2$. 
Between the right and left
sides of the glassy transition region there is a difference 
$\Delta C=Nc_2(1-1/\Q)$.

Furtheron it will become clear that
$m(t)-m_{eq}(T_e)$ remains zero upon cooling at fixed field $H$.
Therefore one has
\BEQ m=m_{eq}(T_e)=\frac{H}{K}-\frac{H}{K^2}\e=\frac{H}{K^2}\,
(\sqrt{K^2+\frac{T_e^2}{4}}-\frac{T_e}{2}) \EEQ
This yields a magnetizability
\BEQ \alpha=\frac{H}{2K^2}(1-\frac{T_e}{\sqrt{4K^2+{T_e^2}}})
\frac{\d T_e}{\d T}\EEQ
and, since $K^2=\Gamma^2+H^2$, a susceptibility $\chi=\chi_{11}$
of the form (\ref{flucts=}), with
\BEQ\label{chiantwfl}
 \chi^{\rm fluct}
=\frac{\Gamma^2+T_e^2/4+wT_e/2}{w(w+T_e/2)^2}
\approx 
\frac{\Gamma^2}{K^3}-\frac{(\Gamma^2-H^2)T_e}{2K^4},
\EEQ
where $w=\sqrt{K^2+T^2_e/4}$, and 
\BEQ \label{chiantcf}
\chi^{\rm conf}=-\frac{H}{w(2w+T_e)}\frac{\p T_e}{\p H}
\approx -\frac{H}{2K^2}(1-\frac{T_e}{2K})\frac{\p T_e}{\p H}\EEQ

Around the glassy transition there occur smeared discontinuities 
in the apparent specific heat, magnetizability and susceptibility
\BEA \label{dCh=}
\Delta c&=&c_2(1-\frac{\p T_e}{\p T}\fix_H) \\
\label{dah=} \Delta \alpha&=&\frac{H}{K^2}c_2 
(1-\frac{\p T_e}{\p T}\fix_H) \\
\label{dchh=}
\Delta\chi&=&\frac{H}{K^2}c_2 \frac{\p T_e}{\p H}\fix_T 
\EEA
These results and eq. (\ref{dTedT=1}) allow us to verify 
the first Ehrenfest relation (\ref{Ehren1H}).

From the identity $\I=S_{eq}(T_e)$ and the expression (\ref{SeqS=})
we obtain
\BEQ
\frac{\p \I}{\p T_e}\fix_{T,H}=\frac{C_2}{T_e}, 
\qquad \frac{\p \I}{\p H}\fix_{T,T_e}=-\frac{HC_2}{K^2} 
\EEQ
We can now consider the modified second Ehrenfest relation 
(\ref{modEhren2H}). Due to eqs. (\ref{dCh=}) and (\ref{dah=}) 
it takes the form
\BEQ \frac{c_2}{T_g}(1-\frac{\p T_e}{\p T}\fix_H)=
\frac{Hc_2}{K^2} (1-\frac{\p T_e}{\p T}\fix_H)\frac{\d H_g}{\d T}
+(1-\frac{\p T_e}{\p T}\fix_H)
(\frac{c_2}{T_g}\,\frac{\p T_e}{\p T}\fix_H
-\frac{Hc_2}{K^2} \frac{\d H_g}{\d T}+
\frac{c_2}{T_g}\,\frac{\p T_e}{\p H}\fix_T\frac{\d H_g}{\d T})
\EEQ
After dividing out the common factor $1-\p T_e/\p T$
and eliminating the remaining $\p T_e/\p T$ by use of the relation
(\ref{dTedT=1}),  it is seen that terms with and without $\d H_g/\d T$
cancel separately. This implies that the modified 
second Ehrenfest relation is satisfied for any value of that
parameter, as was to be expected.

Using (\ref{Pip=}) and the relation $T_g\approx A/\ln t_g$, 
the Prigogine-Defay ratio can now be expressed as
\BEQ\label{PipS=}
\Pi=\frac{\Delta C}{T N\Delta\alpha}\,\frac{ \d T_g}{\d H}
=\frac{K^2}{HT_g}\,\frac{\d T_g}{\d H}
=1-\frac{K^2}{H\ln t_g}\,\frac{\d \ln t_g}{\d H}
\EEQ
Contrary to what was long believed, 
the condition $\Pi<1$ is easily met. Indeed, in case that 
 $\Q$ is fixed, 
we may still choose the glassy transition line $T_g(H)$, or,
equivalently the glassy transition timescale $t_g(H)$. 
Values $\Pi<1$ thus occur when $\d t_g/\d H>0$, so in half of 
the sets of smoothly related cooling sequences. 
This analysis confirms our general argument that the Prigogine-Defay
ratio can take any value between zero and infinity, and perhaps
even negative values.

\subsection{Aging regime  and its thermodynamics}
For temperatures in the aging regime we have, very analoguous to
(\ref{Tedot}), 
\BEQ \label{TedotS}
\frac{\d (e^{\beta_eB})}{\d t}=
\frac{2K^2}{(K-\e)^2}
\sqrt{\frac{BT_e}{\pi}}
(1-r^2)
\EEQ
where we again use the short hand $r={T}/({2T_e-T})$.
To leading order this relation even reduces to eq. (\ref{Tedot}), the 
only change being $\exp(\beta_eA)\to\exp(\beta_eB)
\approx \exp(\beta_eA+A/2K)$.
Therefore to leading order $T_e$ again follows from
eq. (\ref{betae=}), with $A\to A+AT_e/2K$.

Let us stress that we donot consider the regime $\Delta\gg 1$, where
a non-universal regime $1\ll \ln t \ll \Delta^2$ would occur.
This is the subject of a recent work on a related model
with fast and slow processes, in which a Kauzmann transition occurs
~\cite{NVFmodel}.

In next subsection it is made clear that $\mu_1=m_{eq}-m$ is
exponentially small in $T_e$. Therefore the magnetization very closely
follows its quasi-equilibrium value set by $T_e$.

We can now check the thermodynamics.
It holds that 
\BEA \label{Ut=}U&=&N(-K+\frac{T_e}{2}-\frac{T_e^2}{8K})\\
\label{Mt=}
M&=&N(\frac{H}{K}-\frac{HT_e}{2K^2}+\frac{HT_e^2}{8K^2})\\
\label{It=}
\I&=&N(\frac{1}{2}\ln\frac{T_e}{K}+\frac{1}{2}-\frac{T_e}{4K})
\EEA
The first law (\ref{dU=}) with $\S=0$ reduces to $\d U=T_e\d\I-M\d H$.
At constant $H$ it holds because one has replaced $T\to T_e$ in 
energy and entropy. Then one can take the difference between 
aging experiments at two nearby $H$'s. The relation 
remains valid because $M$ is essentially equal to its
quasi-equilibrium value $M_{eq}(T_e,H)$.  

In the modified Maxwell relation (\ref{modMaxH}) the terms without
explicit factor $T$ cancel because of the first law 
(\ref{dU=}) with $\S=0$. The terms with explicit factor $T$ 
cancel separately, because of the quasi-equilibrium relation 
$\p\I/\p H|_{T,T_e}=-\p M/\p T_e|_{T,H}$ that follows from (\ref{dF=}).
Such a pairwise cancellation would, of course, not occur in 
less simple models. Indeed, in a glassy
model with directed polymers ~\cite{Ndirpol} the modified Maxwell
is also satisfied, but in a non-trivial manner~\cite{NEhren}.

\subsection{Changing the external field and the need for an effective field}

If also field $H$ is changed in time, then we have to be more careful.
The Monte Carlo dynamics now leads to 
\BEQ \dot m_a=\int \d x W(\beta\tilde x)\overline y_ap(x|\e)\EEQ
where, again, $m_1=m$, $m_2=m_s$ and $\tilde x=x-\dot H m$.
We shall be interested in cases with logaritmically slow $H$, 
implying $\dot H\sim 1/t$, where $\tilde x\approx x$.

From the definitions (\ref{e=mu1=})
 we derive for the evolution of $\e$ and $\mu_1$
\BEQ \dot \e=\dot H(\frac{H}{K}-m)+\int \dx W(\beta\tilde x)x
p(x|\e)=\dot H(\frac{H}{K^2}\e+\mu_1)+\int \dx W(\beta\tilde x)xp(x|\e)
\EEQ
\BEQ \dot \mu_1=-\dot H(-\frac{\Gamma^2}{K^3}+\e\frac{\Gamma^2-H^2}{K^4}+
(\frac{H}{K}-m)\frac{H}{K^2})-f\mu_1
\approx \dot H\frac{\Gamma^2}{K^3}-f\mu_1
\EEQ
So far we considered cooling at constant field. Then eq. (\ref{ht=})
says that $\mu_1(t)=\mu_1(t_0)h(t_0)/h(t)$ decays as a powerlaw, namely
$1/\sqrt{t}$ for $T>0$ and $1/t$ when $T=0$. Both behaviors are
exponentially small in $T_e$, and much smaller than 
subleading powers of $T_e$ that were neglected already.
So to the accuracy considered we can set $M(t,H)$ equal to $M_{eq}(T_e,H)$.

When the field is slowly changed in the course of time, the leading behavior
of $\mu_1$ is
\BEQ \label{Hdotconst}
\mu_1=\frac{\dot H\Gamma^2}{K^3f} \EEQ
As long as $|\mu_1|\ll HT_e/K^2$ the main change of $M(t)$ is still 
expressed by $M_{eq}(T_e(t),H(t))$. Since $f\sim (1+r)/t$, 
this condition can be written as
\BEQ \left|\frac{\partial H}{\p \ln t}\right|
\ll \frac{HKT_e(1+r)}{\Gamma^2}\quad
\to \quad \left|\frac{\partial H}{\p T_e}\right|
\ll \frac{AHK(1+r)}{\Gamma^2T_e}\quad
\EEQ
This condition is reasonable, and easily satisified near 
$T_e=0$.

When condition (\ref{Hdotconst}) is not fulfilled, it is
not possible to describe $U(t)$ and $M(t)$ by $T_e(t)$ alone. 
One needs a second effective variable, namely the effective field 
$H_e(t)=H(t)+\delta H_e(t)$.
Setting $K_e=\sqrt{\Gamma^2+H_e^2}$, we then have from
quasi-equilibrium formulae at $(T_e,H_e)$
\BEQ u=-K_e+\frac{T_e}{2}\approx -K-\frac{H}{K}\delta
H_e+\frac{T_e}{2}\EEQ
\BEQ m=\frac{H_e}{K_e}(1-\frac{T_e}{2K_e})\approx \frac{H}{K}
+\frac{\Gamma^2}{K^3}\delta H_e-\frac{H}{2K^2}T_e
\EEQ
From the definitions (\ref{e=mu1=})
we can make the identifications
\BEQ \e=\frac{T_e}{2}-\frac{H}{K}\delta H_e;\qquad
\mu_1=-\frac{\Gamma^2-H^2}{K^3}\delta H_e
\EEQ
or their inversion
\BEQ T_e=2\e-\frac{2HK^2}{\Gamma^2-H^2}\mu_1;\qquad
\delta H_e=-\frac{K^3}{\Gamma^2-H^2}\mu_1 \EEQ
One can now consider any class of fields that change
logarithmically slowly in time.
After solving the dynamics $T_e$ and $H_e$ follow.
One can then also study fluctuations and the fluctuation-dissipation
relation for this more general case, and look for universal
behaviors. This matter is the subject of currentresearch
~\cite{NVFmodel}, which falls outside the scope of the present paper.

\subsection{Fluctuations and fluctuation-dissipation relation}
The energy fluctuations still follow from (\ref{dDE2dt=}), and
depend on $H$ only through $K$. They are, to leading order, 
again given by (\ref{explEfluct}),
\BEQ \label{explEfluctS}
\frac{1}{N}\,<\! \delta \H^2\!>=C_{\e\e}(t,t)=
\frac{T_e^3(t)}{A}\,
\frac{1+r^3}{1-r^2}\EEQ
This allows to solve the cross-fluctuations from (\ref{EMcorr}),
\BEQ C_{\e m}(t,t)=-\frac{HT_e^3(t)}{AK^2}\,
\frac{1+r^3}{(1-r)(2-r)}\EEQ
Finally, up to corrections of order $T_e^2$,
the $M$-fluctuations satisfy an equation very similar to
(\ref{dM^2osc}). This yields
\BEQ C_{mm}(t,t)=\frac{\Gamma^2 T_e(t)}{K^3}(1-\frac{T_e}{2K})+\O(T_e^3),\EEQ
where we notice that terms of relative order $T_e/A$ have cancelled
on both sides of the equality sign. 
Considering the fluctuations in $M_1=M$, $M_2=M_s$, we have
\BEQ\label{Cabtt=}
 C_{ab}(t,t)=\left(\frac{-H}{\Gamma}\right)^{a+b-2}
\frac{\Gamma^2 T_e(t)}
{K^3}\,(1-\frac{T_e}{2K})+\O(T_e^3)\EEQ
The difference with the oscillator model of previous section, is that
all four terms $(a,b=1,2)$ now have corrections of relative order
 $T_e^2/A^2$, that  will be neglected from now on. They decay slower than
the terms we keep, but  we are not interested in the model-dependent 
very long-time regime, where they dominate.
 
The time-dependence of the $C_{mm}(t,t')$ follows from (\ref{dCmmdt=}).
In the interesting, not-very-asymptotic regime the term $C_{\e m}$
can be neglected. This implies finally that
\BEQ C_{ab}(t,t')=C_{ab}(t',t')\frac{h(t')}{h(t)} \EEQ
which involves $h(t)$ defined as in eq. (\ref{ht=}), with,
very analogous to (\ref{f=}),
\BEQ f(t)=\int_\minfty^\infty \d x W(\beta x)\frac{4A-x(1-\e/K)}
{2\e(t)(1-\e/(2K)}p(x|\e(t))={8A}p(0|\e)(1+r)
(1+\frac{T_e}{2K}-\frac{r^2T_e}{2A})
\EEQ

The equal-time correlators take the value
\BEQ G_{ab}(t^+,t)=
\frac{8A}{K^3}\,\left(\frac{-H}{\Gamma}\right)^{a+b-2}
p(0|\e)\,(1-\frac{r^2T_e}{2A})
\EEQ
Both $C_{ab}(t,t')$ and 
$G_{ab}(t,t')$ have a time dependence 
$h(t)/h(t')$.
The fluctuation-dissipation relation (\ref{FDR=}) again holds
with the same effective temperature (\ref{tildeTe=})
as in the oscillator model,  
\BEQ\label{tildeTeS=}
\tilde T_e(t)=T_e(t)+\frac{\dot T_e(t)}{f(t)}
=T_e-\frac{T_e^2}{A}\,\frac{2(T_e-T)}{2T_e-T}
\EEQ
In deriving this result we noticed that  terms of relative order 
$T_e/K$, as appearing in eq. (\ref{Cabtt=}), have cancelled.
We can now redo the consistency check of (\ref{Mabhelp14})
and verify that, up to relative order $T_e^2$,
\BEQ \label{5.45}
\int_0^t \d t'G_{mm}(t,t')=\beta_eC_{mm}(t,t)=
\frac{\Gamma^2}{K^3}
(1-\frac{T_e}{2K})
\EEQ
Using that $T_e\approx \Delta^2 K/(8\ln t)$, we also find
\BEQ \frac{\p T_e}{\p H}\fix_{T,t}=\frac{\Delta^2 H}{8K\ln t}=
\frac{HT_e}{K^2} \EEQ
As mentioned before, this is one order of magnitude smaller than
(\ref{5.45}).
The relation (\ref{Mt=}) now implies $\p m/\p T_e|_{T,H}=-H/2K^2$, so
that eq. (\ref{chifluct}) becomes
\BEQ \chi_{mm}^{\rm fluct}=0+\frac{\Gamma^2}{K^3}
(1-\frac{T_e}{2K})-(-\frac{H}{2K^2})\frac{HT_e}{K^2} \EEQ
In view of the prediction (\ref{chiantwfl}), 
this is the desired answer to the considered order.

Also the energy correlation- and response function are essentially 
the same as in the oscillator model. This implies in particular
the fluctuation-dissipation relation for energy fluctuations
of the non-universal form (\ref{Teeefdef}), (\ref{Teeef=}).

Notice that when there are no random fields, $\Gamma=0\,\to\,K=H$
the energy and the magnetization are proportional to each other,
viz. $E=-HM$.
For comparison with realistic glassy systems, the model becomes too poor. 
The above formula for the magnetization seize to hold when
$\Gamma<T_e$. For $\Gamma\to 0$ one finds the magnetization
correlations from the energy correlations. One then finds
the relations  (\ref{Teeefdef}), (\ref{Teeef=}) both for energy and
magnetization.

\section{Discussion}
\setcounter{equation}{0}\setcounter{figure}{0} 
\renewcommand{\thesection}{\arabic{section}.}

In this paper we consider the question whether the
glass transition can be phrased in a thermodynamic framework.
In a series of letters we have given already several arguments in 
favor of this possibility~\cite{Nthermo}~\cite{NEhren}
\cite{Nhammer}. The present, admittedly long, paper is meant to 
explain enough details of this approach to make the picture and its
various steps and assumptions transparant. We do this by
working out in detail two simple models, that, in our feeling, 
are closer to reality than mean field spin glasses.
We have pointed out there that a minimal 
thermodynamic description needs one more parameter to describe the
situation, which could be the age of the glass, or the cooling rate
at which it was formed. We have discussed that for thermodynamics 
a more useful variable is the effective or fictive temperature $T_e$,
introduced half a century ago by Tool~\cite{Tool}.
In this paper we notice that the basic result for the change
of heat in a glassy system,
\BEQ \dbarrm Q \le T\d\S+T_e\d\I, \EEQ
immediately leads to a specific heat $C_p=C_1+C_2\p T_e/\p T$,
that was Tool's starting point for the analysis in the glass formation
region. In section IV.1 we continue along his lines by studying in
detail a  certain non-linear cooling traject, 
proposed recently by us in ref.~\cite{Nhammer}.
This cooling scheme is applicable to any glass forming substance
and expected to give universal scaling curves of
$\p T_e/\p T$ in the glass formation region, independent of
the material considered, provided that the 
glass transition region is narrow. 
It would be most interesting to test this idea on a realistic
glass forming liquid. One should first determine,
for once and for all, the equilibrium timescale $\tau_{eq}(T)$
and then do  glass experiments of the type (\ref{coolTQ}) 
within the considered range. It contains two parameters: 
the glass formation temperature $T_g$, where the cooling
timescale becomes comparable to the equilibration timescale,
and a ``speed'' parameter $\Q$. The resulting form for
$\p T_e/\p T$ lies, after rescaling the width,
on  a universal scaling curve, that only depends
on $\Q$.  
 
We have here worked out the situation where the effective
temperature shows up as extra variable, though
in principle it might be needed to consider as many
effective parameters as there are macroscopic observables.
We had already shortly considered the Ising chain with Glauber
dynamics. In that model the non-equilibrium energy at zero-field
 can be described by introducing the effective temperature. Its 
definition then coincides by equating the $\tau_{eq}(T_e)$ with $t$.
The behavior at non-zero field appears to be non-universal
~\cite{Nthermo}. More or less the same happens in the
backgammon model, for which the dynamics at zero
field has been partly solved~\cite{FranzRitort}\cite{GodrecheLuck}. 
One could couple the system to a particle bath, and the chemical potential 
would play the role of an external field. So far this case remains to
be worked out.
We have, therefore, focussed on very simple, exactly
solvabel models, namely the Bonilla-Padilla-Ritort model
of Monte Carlo dynamics of uncoupled harmonic harmonic oscillators
~\cite{BPR}, and our recent model of Monte Carlo dynamics of 
uncoupled spherical spins in a random field~\cite{Nhammer}.
At low temperature both models have an Arrhenius law for the 
equilibrium timescale. Upon cooling from high temperatures,
they will sooner or later fall out of equilibrium.
Though it may come as a surprise for some readers,
we have shown that these oversimplified models with their unrealistic
dynamics still share in their off-equilibrium phase
universal properties of realistic glasses and models for glasses.

A description with only the effective temperature applies
whenever the volume of the glass forming liquid (or the 
magnetization of the glassy magnet) is close to its quasi-equilibrium
value set by the effective temperature obtained from fitting the energy.
If this condition is fulfilled, the old objections against a
thermodynamic description of the glassy phase can be inspected 
in detail. We have stressed that the most fundamental paradox,
namely violation of the first Ehrenfest relation, is merely based
on the misleading expectation that there is one, ideal value for the
compressibility. This notion has arosen from 
 equilibrium considerations for the glassy state. They donot apply
by definition, and have hindered progress until our recent works
in this field. Indeed, from the knowledge of spin glasses,
or from the solution of the present models (where $\chi_{ZFC}=0$), 
we know that the compressibility or the susceptibility
can take a broad range of values immediately below the
glassy transition. This means that no alternative determination
is allowed, removing immediately the whole paradox: though half
a century of research led to the general belief that the first
Ehrenfest relation is always violated, it is actually 
satisfied automatically~\cite{NEhren}. This point is 
underlined in figure 2.1, where we present a 3d-plot of 
data for the glass transition in atactic polystyrene,
collected in careful experiments by Rehage and Oels~\cite{RehageOels}.
Though these authors claim that the first Ehrenfest relation is
violated, and then continue to investigate a modified version,
we explain that it is satisfied.

We have also pointed out that for glass forming liquids 
the Maxwell relation between $\p U/\p p$
and $\p V/\p T$  is modified,
which is not so surprising in view that equilibrium is not reached.
The second Ehrenfest relation is then also modified, 
since it relies on the Maxwell relation.  
This fact implies that the Prigogine-Defay ratio can indeed be
different from unity. We should recall that Davies and Jones
~\cite{DaviesJones} showed that $\Pi\ge 1$, while DiMarzio found that a 
deeper analysis of their equations leads to $\Pi=1$~\cite{DiMarzio}.
Both approaches, however, are based on the assumption that at
the glass transition an unspecified number of order parmeters
freezes in, an assumption that was often made in the sixties and
seventies. Such assumptions are invalid, however.
What happens at the glass transition is that certain slow modes
fall out of equilibrium, but on longer scales they may reach
equilibrium again, even though other modes may then have fallen
out of equilibrium. The upshot of this is that the Prigogine-Defay
ratio can be different from unity. In contradiction to the 
standard belief, it can also be less than unity. 
We have pointed out before ~\cite{NEhren} that this already occurs in 
 experiments on atactic polystyrene~\cite{RehageOels}, though this
was long not recognized.

It has been the important contribution of statistical physics to
relate temporal fluctuations in macroscopic observables to their
averages, the most known relation being $C=\beta^2 <\!\delta H^2\!>$.
It is natural to investigate whether such relations have some 
universal-looking generalization in simple models for glasses.
We have found that this is indeed the case for fluctuations
in observables coupled to external fields~\cite{Nhammer}, 
see Section 2.5. These equations have been guessed with an eye on 
results from present models, in combination with
 some standard arguments on the short-time contributions. 

Our progress was initially hindered by the fact
that such general formula appear not to hold for energy fluctuations.
For the ferromagnetic Ising chain aging at $T=0$ from a random
initial condition, we already realized that
at zero field the energy defines an effective temperature
$T_e=2J\ln 4\pi t$,  that coincides,
to leading order, with the one following from equating
the time scale with time, viz. $\tau_{eq}(T_e)=t$~\cite{Nthermo}.
The energy correlations can be calculated from the
Derrida-Zeitak spatial correlation function for the non-meeting
of two random walkers on a line. Indeed, the Ising chain
is mapped to random walking of interfaces by setting
$s_{i-1/2}s_{i+1/2}=1-2\rho_i$ with $\rho_i=1$ if an interface
is present at $i$, and zero else. It follows that $U/N=J(-1+2\rho)$ where
$\rho=1/\xi$ is the average density of interfaces, with $\xi=
\sqrt{4\pi t}\equiv\exp(2\beta_eJ)$ the correlation length.
It holds that
\BEQ \frac{1}{N}\langle\delta U^2\rangle=4J^2(\rho-\rho^2)+
8J^2\sum_{i>0} (C_{i,0}-\rho^2) \EEQ
where $C_{ij}=\langle\rho_i\rho_j\rangle=C(r_{ij})$ is the correlation
function, given by eq. (59) with $q=2$ in a paper by Derrida and Zeitak
{}~\cite{DerridaZeitak},
\BEQ C(r)=\rho^2(1-e^{-2z^2}+2ze^{-z^2}\int_z^\infty e^{-u^2}\d u),\EEQ
where $z=\sqrt{\pi}\,\rho\,r$. One thus finds
\BEQ \frac{1}{N}\langle \delta U^2\rangle=4J^2\rho(3-2\sqrt{2})
\EEQ
This differs from the naively  expected quasi-equilibrium
result $T_e^2\d U/\d T_e=
4J^2\rho$ by a numerical prefactor $3-2\sqrt{2}=0.17157288$.
We were informed by J.M. Luck that in the backgammon model
a similar phenomenon occurs: the energy correlations are a factor two
smaller than naively expected~\cite{JMLprivatecom}.
In the models of the present work the energy fluctuations are even smaller
by an order of magnitude in $T_e/A$. These explicit examples show that
there cannot exist a simple quasi-universal formula relating energy
 fluctuations with the specific heat.
The underlying reason hereto is that in the energy the leading
fluctuations from different terms already cancel, leaving
model-dependent, subleading effects only.

After completion of the original manuscript, mr. L. Leuzzi verified that 
both in the harmonic oscillator model and in the spherical spin model 
there holds the following relation between the specific heat 
and the energy fluctuations:
\BEA\label{dUdT=fluct}
\frac{\p U}{\p T}\fix_H=\frac{1}{T\,T_e^{(\e\e)}}
\langle\delta \H^2\rangle -
\frac{\p U}{\p T_e}\fix_{H,T}\left(\frac{\d T_e}{\p T}\fix_{H,t}-1\right)
+\frac{\p U}{\p T_e}\fix_{H,T}\left(\frac{\d T_e}{\p T}\fix_{H}-1
\right)\EEA
where $T_e^{(\e\e)}$ is given by eq. (\ref{Teeef=}) for both models.
The first two terms in the right hand side cancel, 
and so do, of course, the $\pm 1$ terms.
In analogy with eq. (\ref{flucts=}), we could interpret this relation as
$C=C^{\rm fluct}+C^{\rm loss}+C^{\rm conf}$, but the factor
$1/T$ in the first term could imply that this attempted 
generalization is 
special to our present oscillator and spin models, as is already
expressed by the fact that $T_e^{(\e\e)}\neq T_e$.
It was also realized that for the cross derivatives $\p U/\p H$ and
$\p M/\p T$ similar relations either  do not occur, or are much more
complicated. The same holds for the fluctuation-dissipation relations
connected to these two quantities.

Nevertheless, fluctuations in observables that couple 
to global external fields, appear to behave in a universal way,
at least to leading and dominant subleading order.
These fluctuations are interrelated with the off-equilibrium
fluctuation-dissipation relation (FDR). Originally observed by
Horner~\cite{Horner1,Horner2,CHS} and then extended by
Cugliandolo and Kurchan~\cite{CuKu}, this has become a 
popular test for glassiness of model systems~\cite{Parisi}
\cite{KPS}
\cite{Sellitto}\cite{KobB}. For the models considered the effective
temperature showing up in the FDR is essentially the same as
the one occurring in thermodynamics.

Absence of a universal
relation between energy fluctuations and the specific heat is
very welcome for gravitating systems, which often have a 
negative specific heat. We indeed showed that the present
approach can immediately be applied to phrase the laws
of black hole dynamics in a non-equilibrium
thermodynamic framework~\cite{Nblackhole}. 
The role of bath temperature is played by the cosmic backgound
temperature, that of the effective temperature by the Hawking
temperature, that of the configurational entropy by Bekenstein's
black hole entropy, while the short times processes 
have no sizeable entropy. These ingredients bring a very close
analogy with the picture discussed here, and are not even disturbed
by the fact that the specific heat has the ``wrong'' sign.
Let us mention, however, that negative specific heats are in no way
limited to gravitation: they also occur
in the present, extensive solid state models, when one
heats up the system in the glassy phase, as expressed by the
negative values of $\d T_e/\d T=2C$ in figures 4.1-4.4.

Our picture for thermodynamics of the glassy state thus connects
macroscopic observables via the first and second law,
and relates their derivatives with respect to external fields
with their fluctuations, thereby embedding the
FDR-effective temperature in a larger thermodynamic framework.
It is expected to be valid for a yet unknown class of glassy systems.
Let us close this discussion by mentioning that very recently 
numerical data in the glassy phase of a binary Lennard-Jones system 
were interpreted in terms of an effective temperature, 
that dominates the (short-time) vibrational properties, 
in full harmony with the picture proposed above~\cite{KobSciorTart}.

\acknowledgments
The author thanks 
E. Hennes,
J.M. Luck, 
F.G. Padilla,
G. Parisi, and 
F. Ritort
for discussion, and E. Hennes for allowing to reproduce figures
4.1 - 4.4. Special thanks for moral support are due to
W.A. van Leeuwen and G. H. Wegdam. 
Last but not least, it is a pleasure to  thank
L. Leuzzi for checking all derivations of the paper, and for
allowing us to discuss his findings concerning eq. 
(\ref{dUdT=fluct}).



\references
\bibitem{GibbsDiMarzio}
J.H. Gibbs and E.A. DiMarzio, J. Chem. Phys. {\bf 28} (1958) 373
\bibitem{AdamGibbs}
G. Adam and J.H. Gibbs, J. Chem. Phys. {\bf 43} (1965) 139
\bibitem{DiMarzio1981} E.A. DiMarzio, Ann. NY Acad. Sci. {\bf 371} (1981) 1
\bibitem{Angell} C.A. Angell, Science {\bf 267} (1995) 1924
\bibitem{RehageOels} G. Rehage and H.J. Oels,
High Temperatures-High Pressures {\bf 9} (1977) 545
\bibitem{NEhren} Th.M. Nieuwenhuizen, Phys. Rev. Lett. {\bf 79}(1997) 1317
\bibitem{Nthermo} Th.M. Nieuwenhuizen, J. Phys. A {\bf 31} (1998) L201
\bibitem{Binder} Wolfgardt M., Baschnagel J., Paul W., and Binder K.,
Phys. Rev. E {\bf 54} 1535 (1996)
\bibitem{NVFmodel} Th.M. Nieuwenhuizen, preprint (1998)
\bibitem{Nhammer} 
Th.M. Nieuwenhuizen, Phys. Rev. Lett.  {\bf 80} (1998) 5580
\bibitem{Horner1} H. Horner, Z. Phys. B {\bf 86} (1992) 291
\bibitem{Horner2} H. Horner, Z. Phys. B {\bf 87} (1992) 371
\bibitem{CHS} A. Crisanti, H. Horner, and H.J. Sommers, 
Z. Phys. B {\bf 92} (1993) 257
\bibitem{CuKu} L. F. Cugliandolo and J. Kurchan, Phys. Rev. Lett.
{\bf 71} (1993) 173
\bibitem{BCKM}
J.P. Bouchaud, L. F. Cugliandolo, J. Kurchan, and M. M\'ezard,
Physica A {\bf 226} (1996) 243 

\bibitem{Nmaxmin} Th.M. Nieuwenhuizen,
 Phys. Rev. Lett. {\bf 74} (1995) 3463; cond-mat/9504059


\bibitem{Ndirpol} Th.M. Nieuwenhuizen, Phys. Rev. Lett.  {\bf 78}
(1997) 3491

\bibitem{FranzRitort} S. Franz and F. Ritort,
J. Phys. A Lett. {\bf 30} (1997) L359

\bibitem{GodrecheLuck} C. Godreche and J.M. Luck, 
J. Phys. A {\bf 30} (1997) 6245
\bibitem{BPR} L.L. Bonilla, F.G. Padilla, and F. Ritort,
Physica A {\bf 250 }(1998) 315

\bibitem{KPS} J. Kurchan, L.  Peliti, M. Sellitto, 
Europhys. Lett. {\bf 39} (1997) 365 
\bibitem{Sellitto} M. Sellitto, cond-mat/9804168

\bibitem{Hammann} F. Lefloch, J. Hammann, M. Ocio, and E. Vincent,
Europhys. Lett. {\bf 18} (1992) 647

\bibitem{LDAHDAice} 
O. Mishima, L.D. Calvert, and  E. Whalley, Nature {\bf 314}
(1985) 76

\bibitem{HES}
O. Mishima and H.E. Stanley, Nature {\bf 392} (1998) 164


\bibitem{Mottishaw} P. Mottishaw, Europhys. Lett. {\bf 1} (1986) 409 

\bibitem{Goldschmidt}  Y. Y. Goldschmidt, Phys. Rev. B {\bf 41}(1990) 4858 
\bibitem{ThirumDobrov} V. Dobrosavljevic and D. Thirumalai, 
J. Phys. A {\bf 22}, (1990) L767 

\bibitem{NieuwRitort} Th.M. Nieuwenhuizen and F. Ritort,
Physica A {\bf 250} (1998) 8

\bibitem{Tool} A.Q. Tool, J. Am. Ceram. Soc. {\bf 29} (1946) 240.

\bibitem{DaviesJones} R.O. Davies and G.O. Jones,
 Adv. Phys. {\bf 2} (1953) 370

\bibitem{Inamina} I. Hodge, Science {\bf 267} (1996) 1945

\bibitem{Goldstein}
M. Goldstein, J. Phys. Chem. {\bf 77} (1973) 667

\bibitem{Jaeckle}
J. J\"ackle J. Phys: Condens. Matter {\bf 1} (1989) 267; eq, (9)

\bibitem{Mydoshboek} J.A. Mydosh,
{\it Spin glasses: an experimental introduction}
(Taylor and Francis, London, 1993)

\bibitem{McKenna} G.B. McKenna,
in {\it Comprehensive Polymer Science, Vol. 2}, eds. C. Booth and C. Price,
Pergamon, Oxford, (1989), pp 311


\bibitem{DiMarzio} E.A. DiMarzio, J. Appl. Phys. {\bf 45} (1974) 4143

\bibitem{Nblackhole} Th.M. Nieuwenhuizen, 
Phys. Rev. Lett. {\bf 81} (1998) 2201-2204


\bibitem{EHennes}  E. Hennes, private communication (1998)

\bibitem{DerridaZeitak} B. Derrida and R. Zeitak, 
Phys. Rev. E {\bf 54} (1996) 2513 
\bibitem{JMLprivatecom} J.M. Luck, private communication (1997)
\bibitem{Parisi} G. Parisi,
Phys. Rev. Lett. {\bf 79} (1997) 3660




\bibitem{KobB} J.L. Barrat and W. Kob, 
{\it Fluctuation dissipation ratio in an aging Lennard-Jones glass},
cond-mat/9806027 

\bibitem{KobSciorTart} W. Kob, F. Sciorti
{\it Aging as dynamics in configuration sp
ace},
cond-mat/9905090
\end{document}